\documentclass[
  aps,
  prx,
  nofootinbib,
  reprint,
  floatfix,
  superscriptaddress,
  twocolumn,
  10pt,
  a4paper,
]{revtex4-2}
\usepackage[utf8]{inputenc}
\usepackage[T1]{fontenc}

\usepackage[paperwidth=210mm,paperheight=297mm,centering,hmargin=2cm,vmargin=2cm]{geometry}

\usepackage{amsfonts,amsmath,amssymb,mathtools}

\usepackage{graphicx}

\usepackage{physics}
\usepackage{bm}

\usepackage[USenglish]{babel}
\usepackage{hyperref}%
\usepackage[capitalize]{cleveref}
\usepackage[dvipsnames]{xcolor}

\renewcommand{\i}{\mathrm{i}}

\newcommand{\pdftitle}{Universal scaling framework for parameterized quantum evolutions at criticality}

\newcommand{\pdfkeywords}{Variational Quantum Algorithms, Renormalization Group, Universal Scaling Analysis}

\hypersetup{%
  colorlinks = true,%
  citecolor  = BrickRed,%
  linkcolor  = RoyalBlue, %RoyalBlue,
  urlcolor   = RedViolet, %RoyalBlue,
  unicode, %
  pdftitle=\pdftitle,
  pdfsubject={Preprint version},
  pdfauthor={Jan~T.~Schneider, Cristian~Tabares, Alejandro~González-Tudela, and Luca~Tagliacozzo},
  pdfkeywords={\pdfkeywords},
  pdfduplex = DuplexFlipLongEdge,
  pdflang = en,
}%

\usepackage[
  activate={true,nocompatibility}, % Activate protrusion and expansion
  final,                           % Enable microtype even in draft mode
  tracking=true,
  kerning=true,
  spacing=true,
  factor=1100,                     % Add 10% to the protrusion amount
  stretch=20,                      % Reduce stretchability (default 20)
  shrink=20                        % Reduce shrinkability (default 20)
]{microtype}
\microtypecontext{spacing=nonfrench}

\begin{document}

%Title of paper
\title{\pdftitle}

\newcommand{\QUARC}{Quantum Advanced Research Center (QuARC), CSIC, Calle Serrano 113b, 28006 Madrid, Spain}
\newcommand{\IFF}{Institute of Fundamental Physics (IFF), CSIC, Calle Serrano 113b, 28006 Madrid, Spain}

\author{Jan~T.~Schneider}
\email{jan.schneider@csic.es}
\affiliation{\QUARC}
\affiliation{\IFF}

\author{Cristian~Tabares}
\affiliation{\QUARC}
\affiliation{\IFF}

\author{Alejandro~González-Tudela}
\email{a.gonzalez.tudela@csic.es}
\affiliation{\QUARC}
\affiliation{\IFF}

\author{Luca~Tagliacozzo}
\email{luca.tagliacozzo@iff.csic.es}
\affiliation{\QUARC}
\affiliation{\IFF}

\date{\today}

\begin{abstract}
Variational ansätze are a cornerstone of quantum many-body physics, providing compact approximations to complex ground states using finite resources.
Recent quantum-technology advances have introduced a new class based on layered parameterized evolutions.
Assessing whether they can represent critical ground states is challenging: correlations span all length scales, while finite circuit depth limits how far they extend.
Building on finite-resource scaling from tensor networks, we assign each ansatz an emergent correlation length $\xi_D$, the longest range over which it faithfully captures critical correlations.
Its growth with refinement parameter $D$, $\xi_D \propto D^\kappa$, defines an exponent $\kappa$ measuring how efficiently an architecture converts resources into long-distance correlations.
Applying this framework to the critical transverse-field Ising model, with $D$ the circuit depth of parameterized evolutions, we compare ansätze with nearest-neighbor and long-range generators, and layers where generators act separately or combined.
All ansätze are compatible with algebraic growth, but fitted exponents range from $\kappa\simeq1$ to $\kappa\simeq3$.
Exponential interactions give the largest exponents, while power-law interactions stay close to nearest-neighbor behavior, showing long-range support alone gives no scaling advantage.
Combined-generator layers generally outperform separable ones, so layer organization matters alongside interaction range.
Finally, a quasiparticle analysis shows finite depth leaves an unresolved window of width $\xi_D^{-1}$ around the low-energy modes responsible for long-distance correlations.
The emergent correlation length thus acts as an infrared resolution scale, providing a benchmark for critical-state preparation and a guide for designing resource-efficient variational architectures.
\end{abstract}

\maketitle

\section{Introduction}

Quantum many-body systems remain a fundamental challenge as the complexity of their description grows exponentially with the number of constituents. Variational ansätze have long provided a practical route to address this problem by approximating ground states within structured and physically motivated families of states. Their success ranges from early mean-field approaches such as Hartree--Fock~\cite{hartree_wave_1928,fock_naherungsmethode_1930}, through correlated-pair constructions~\cite{jastrow_many-body_1955,bardeen_theory_1957}, to modern tensor-network methods~\cite{ranTensorNetworkContractions2020,cirac2021,banuls_tensor_2023,berezutskii_tensor_2025}. Across these approaches, the central task is to capture the relevant correlations of a system with limited resources. This challenge becomes particularly acute at criticality, where correlations extend over all length scales and therefore demand increasingly large resources for accurate representation.

Recent works~\cite{kandala2017,cerezo2021,bravo-prieto2020,jobst2022,tabares2023,lyu_variational_2023,tabares_programming_2026,weiUniversalEfficientHybrid2026,Kokail2019,Joshi2023,andersen_thermalization_2025} have suggested that new families of variational ansätze implemented directly on quantum hardware can approximate critical ground states. These approaches were first introduced as parameterized quantum circuits built from sequences of single- and two-qubit gates~\cite{kandala2017,cerezo2021,bravo-prieto2020,jobst2022}. More recently, hybrid digital--analog strategies~\cite{tabares2023,lyu_variational_2023,tabares_programming_2026,weiUniversalEfficientHybrid2026,Kokail2019,Joshi2023,andersen_thermalization_2025} have broadened this class to parameterized quantum evolutions, in which hardware-native many-body interactions are interleaved with on-site control operations.
Such evolutions are often more experimentally natural than fully digital circuits and can be realized across a wide range of quantum simulation platforms~\cite{islam2011,lanyon2011,blatt2012,schneider2012,bermudez2017,monroe_programmable_2021,chomaz_dipolar_2023,weimer2010,schauss2012,browaeys2016,scholl2021,bluvstein2024,manetsch2024,goban13a,goban15a,hood16a,Samutpraphoot2020,laucht12a,evans2018,Appel2021,tiranov2023,MacHielse2019,Rugar2020,Rugar2021,liu17a,Mirhosseini2018a,Sundaresan2019,Scigliuzzo2022,Zhang2022Simulator,krinner18a,Kwon2022FormationLattice,douglas2015,gonzalez-tudela2015,Hung2016}.
However, existing studies have largely focused on circuit-specific causal-cone structures~\cite{bravo-prieto2020,jobst2022}, finite-size performance~\cite{tabares2023,lyu_variational_2023,tabares_programming_2026,weiUniversalEfficientHybrid2026}, or particular hardware implementations.
Consequently, a general framework to compare the efficiency of different variational architectures at criticality is still lacking.

\begin{figure*}[t]
  \centering
  \includegraphics[width=\linewidth]{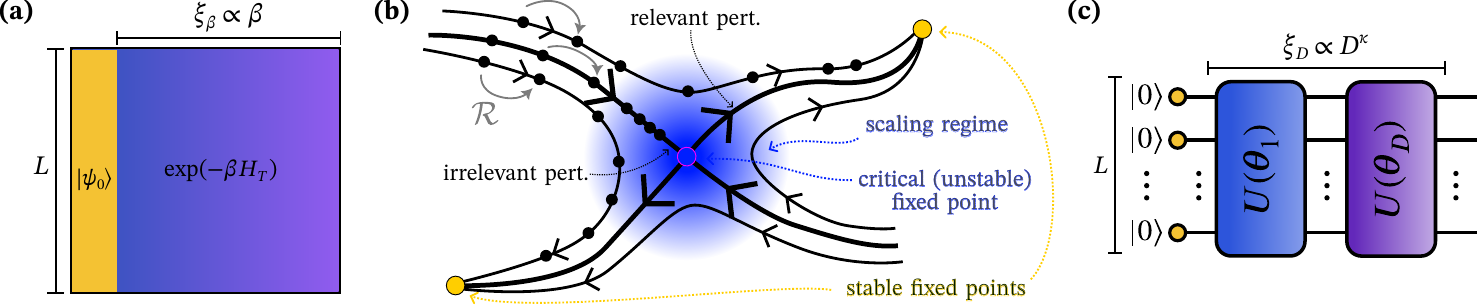}
\caption{\label{fig:partition-circuit-RG}%
\textbf{Conceptual overview of the universal scaling framework.}
The search for the ground state $\ket{E_\mathrm{GS}}$ of a target many-body Hamiltonian $H_T$ can be viewed as a two-dimensional problem, where the system size $L$ competes with a finite resource that limits the accessible correlation length $\xi$.
(a) In the imaginary-time formulation, $\ket{E_\mathrm{GS}}\propto\lim_{\beta\rightarrow\infty}e^{-\beta H_T}\ket{\psi_0}$ is represented by a partition function with spatial extent $L$ and imaginary-time extent $\beta$; stopping the projection at finite $\beta$ induces a finite correlation length $\xi_\beta\propto\beta$.
(b) From the RG perspective, a finite-resource approximation acts as an effective perturbation away from the critical fixed point and induces a finite correlation length $\xi$, while universal scaling behavior emerges in the scaling regime.
(c) Parameterized quantum evolutions provide an analogous $L\times D$ structure, where the circuit depth $D$ induces an emergent correlation length $\xi_D$.
In the scaling regime, observables depend on the finite resource only through $\xi_D$, so the growth law $\xi_D\propto D^\kappa$ provides a common measure of variational efficiency across different ansätze.}
\end{figure*}

In this work, we therefore adopt the renormalization-group-based finite-resource scaling perspective established in tensor-network theory~\cite{nishino_numerical_1996,tagliacozzo2008,pollmann2009,tagliacozzoSimulationTwodimensionalQuantum2009,pirvu2012,evenbly2013,corboz_2018,rader_finite_2018,czarnik_finite_2019,vanhecke_scaling_2022,ueda_tensor_2022} and apply it to parameterized quantum evolutions.
This perspective associates each finite-resource variational ansatz with an emergent correlation length $\xi_D$, as illustrated in \cref{fig:partition-circuit-RG}.
We denote the refinement parameter controlling the available resource by $D$. For the parameterized quantum evolutions considered here, $D$ is the circuit depth. We use the growth of the emergent correlation length,
 \[
\xi_D\propto D^\kappa,
\]
as a quantitative measure of variational efficiency. Because the framework depends only on this emergent infrared scale, the definition and extraction of $\xi_D$ are independent of the microscopic realization of the ansatz, while the exponent $\kappa$ provides a common figure of merit for comparing distinct variational architectures.
We apply this framework to several families of parameterized quantum evolutions targeting the critical transverse-field Ising model. Its free-fermion structure allows all observables used in the optimization and scaling analysis to be evaluated numerically exactly at large system sizes. The ansätze span two design axes: their interaction structure, comparing nearest-neighbor and long-range generators, and their layer organization, comparing separable Hamiltonian Variational Ansatz (HVA) layers with combined evolutions generated by a single exponential.

Within the explored depth and system-size windows, all ansätze are compatible with algebraic growth of $\xi_D$, but with substantially different exponents. Exponential long-range generators yield the largest fitted values of $\kappa$, whereas power-law interactions remain much closer to nearest-neighbor behavior, showing that long-range support alone does not guarantee a scaling advantage. Combined layers also generally yield larger fitted exponents than their separable HVA counterparts, suggesting that the organization of the variational generators matters alongside their interaction range.

Finally, we develop a microscopic interpretation of the emergent correlation length by analyzing the optimized variational states in the fermionic quasiparticle basis of the target Hamiltonian. We show that finite circuit depth leaves an unresolved window around the low-energy gap-closing modes whose width scales as $\xi_D^{-1}$, while modes away from criticality are already faithfully reproduced at shallow depth. This identifies the variational correlation length as an infrared resolution scale and recasts the variational problem as resolving the non-analytic occupation step of the critical ground state. The ansatz-dependent exponent $\kappa$ therefore quantifies how efficiently different variational architectures convert finite resources into long-distance correlations.

The rest of the manuscript is organized as follows. Section~\ref{sec:framework} reviews the finite-resource scaling perspective and the observables used to extract $\xi_D$ from the variational state. Section~\ref{sec:parameterized} defines the parameterized quantum evolutions considered in this work. Section~\ref{sec:application} applies the scaling analysis to the critical TFIM and presents the numerical results together with their microscopic interpretation. Finally, Section~\ref{sec:conclusion} summarizes the implications and outlines directions for future work.

\section{Universal scaling framework}\label{sec:framework}

In this section, we review the universal finite-resource scaling perspective established in tensor-network theory and formulate its application to parameterized quantum evolutions. The guiding idea is to use universality as a common language for finite-resource variational approximations. Near a critical point, RG theory separates emergent long-distance scaling from microscopic details. Here we use this separation to compare how different variational ansätze convert a refinement parameter into an emergent infrared correlation length $\xi_D$. The growth of $\xi_D$ with the refinement parameter then defines an efficiency exponent $\kappa$, which is ansatz dependent but extracted within a common scaling language. This section is organized as follows. In Section~\ref{subsec:background}, we review the RG perspective underlying the framework. In Section~\ref{subsec:extraction}, we show how $\kappa$ can be extracted from physical observables.

\subsection{RG perspective on finite-resource approximations\label{subsec:background}}

A central challenge in quantum many-body physics is the faithful representation of ground states, $\ket{E_\mathrm{GS}}$, of a target many-body Hamiltonian, $H_T$, using finite resources. While variational ansätze provide a controlled way to approximate such states, finite resources generically limit the range of correlations that can be faithfully captured. This limitation becomes particularly severe at criticality, where the gap $\Delta_\mathrm{GS}$ closes, the ground state becomes scale invariant, and correlations extend over arbitrarily large distances.

A useful way to view this problem, illustrated in \cref{fig:partition-circuit-RG}(a), is through the imaginary-time representation of the ground-state projector,
\begin{equation}
  \ketbra{E_\mathrm{GS}}
  =
  \lim_{\beta\to\infty}\rho(\beta)
  =
  \lim_{\beta\to\infty}
  \frac{\exp(-\beta H_T)}{\mathcal{Z}}\,,
  \label{eq:gs_proj}
\end{equation}
with $\mathcal{Z}=\tr[\exp(-\beta H_T)]$. Excited-state contributions are suppressed as $\exp(-\beta\Delta_\mathrm{GS})$, so the closing of the spectral gap $\Delta_\mathrm{GS}=\Delta_\mathrm{GS}(L)$ with system size $L$ controls the inverse temperature $\beta$ required to project onto the ground state. Conversely, stopping the projection at finite $\beta$ limits the range of correlations that can be faithfully reproduced, inducing a finite correlation length $\xi_\beta$~\cite{sachdevQuantumPhaseTransitions2011}.

From the perspective of the RG~\cite{wilson1975,cardy1996}, critical ground states are characterized by the absence of an intrinsic length scale. Their long-distance properties are therefore universal and determined only by the corresponding universality class. Within this picture, a finite-resource approximation can be represented at long distances as an effective perturbation that displaces the system from the critical fixed point~\cite{wilson1975,cardy1996,tagliacozzo2008,pollmann2009,schneiderSelfcongruentPointCritical2025}. Under coarse-graining transformations $\mathcal{R}$, perturbations are classified as \emph{relevant}, which grow under the RG flow, or \emph{irrelevant}, which are progressively suppressed, as illustrated in \cref{fig:partition-circuit-RG}(b). Finite-resource approximations generically induce effective relevant perturbations, giving rise to a finite correlation length $\xi_D$ that sets the largest scale over which the variational state faithfully reproduces the critical correlations.

This perspective provides a natural, ansatz-independent framework for assessing the efficiency of variational approximations.
Let $D$ denote a refinement parameter of the ansatz that systematically improves the approximation.
Analogously to the imaginary-time extent $\beta$, the refinement parameter $D$ induces an effective correlation length $\xi_D$ that characterizes the longest correlations faithfully reproduced by the variational state.
In this work, $D$ corresponds to the circuit depth of parameterized quantum evolutions illustrated in \cref{fig:partition-circuit-RG}(c), but the framework applies equally to other resources such as the bond dimension of tensor-network ansätze. The remaining question is therefore how the induced correlation length $\xi_D$ grows with the refinement parameter $D$.

For sufficiently large $D$, depending on the initial state, the variational state enters the scaling regime. There, the efficiency of the ansatz is determined by how the induced correlation length $\xi_D$ grows with the refinement parameter $D$. If $\xi_D$ grows logarithmically,
\begin{equation}
  \xi_D\sim\log D,
\end{equation}
the ansatz requires exponential resources to resolve long-range correlations and is therefore inefficient. If instead the growth is algebraic,
\begin{equation}
  \xi_D=A_\xi D^\kappa,
  \label{eq:xi-scaling}
\end{equation}
with $A_\xi$ a non-universal constant and $\kappa>0$, the ansatz captures long-range correlations with polynomial resources. Faster-than-algebraic growth would correspond to an even more efficient representation, although such behavior is not expected generically. In this work, we focus on the algebraic regime and use the exponent $\kappa$ as a quantitative figure of merit to compare different variational strategies at criticality.

\subsection{Extracting the scaling exponent from observables\label{subsec:extraction}}

We now explain how the scaling exponent $\kappa$ is extracted in practice. Rather than fitting $\kappa$ directly, the strategy is to first determine the effective correlation length $\xi_D$ induced by the finite variational resources, and then study how it grows with the refinement parameter $D$. We estimate $\xi_D$ independently from both correlation functions and energy errors, providing two complementary determinations of the scaling exponent $\kappa$. The RG arguments reviewed below explain why these independent estimates are expected to agree. We formulate the analysis in one spatial dimension, while keeping the discussion independent of any particular universality class.

The central assumption of RG is that finite variational resources modify the optimized state only through its long-distance properties. Consequently, at sufficiently large distances, the optimized finite-depth state can be described as the ground state of the critical target theory perturbed by the leading relevant operator allowed by the symmetries of the ansatz. Keeping only this leading contribution, we write the effective Hamiltonian as
\begin{align}
  H_D
  =
  H_T
  +
  g_D\int_0^L \Phi(x)\dd x\,,
  \label{eq:effective-hamiltonian}
\end{align}
where $H_T$ is the target Hamiltonian of a critical system, $g_D$ is the effective coupling induced by the finite variational resources, and $\Phi$ is the leading relevant perturbing operator allowed by the symmetries of the ansatz. Such a perturbation drives the system away from criticality, opening a gap and generating a finite correlation length,
\begin{align}
  \xi_D \propto g_D^{-\nu},
  \qquad
  \nu^{-1}=d-\Delta_\Phi,
  \label{eq:xi-perturbation}
\end{align}
where $\Delta_\Phi$ is the scaling dimension of the perturbing field $\Phi$ and $d$ is the spacetime dimension, which in the applications below is $1+1$.

The finite variational correlation length $\xi_D$ and the finite system size $L$ therefore define two competing infrared cutoffs. When $\xi_D\gg L$, the finite system size is the dominant infrared cutoff and the variational state is effectively critical over the entire chain. Conversely, when $\xi_D\ll L$, the induced correlation length $\xi_D$ becomes the dominant infrared cutoff, limiting the range over which critical correlations are faithfully reproduced. The purpose of the following analysis is to make this competition operational. Only in the finite-depth regime $L\gg\xi_D$, where $\xi_D$ rather than $L$ cuts off the RG flow, do universal scaling forms provide a clean readout of the variational correlation length and therefore of how efficiently variational resources are converted into critical correlations.

In the one-dimensional systems considered here, the conformal field theory of the corresponding universality class predicts universal scaling forms in which long-distance observables depend on the ratio $L/\xi_D$. As an illustration, an order parameter obeys the finite-size scaling form~\cite{privman_universal_1984,cardy1996}
\begin{equation}
  m(L,D)
  =
  L^{-\beta_c/\nu}
  \mathcal{M}\left(\frac{L}{\xi_D}\right),
  \label{eq:fcs-ansatz}
\end{equation}
where $\beta_c$ is the order-parameter critical exponent and $\mathcal{M}$ is a universal scaling function up to non-universal normalizations.

For the parameterized quantum evolutions studied here, however, both the parameterized evolution and the initial state preserve the symmetries of the target Hamiltonian. Consequently, symmetry-breaking order parameters remain identically zero and cannot be used to extract $\xi_D$. We therefore focus on two complementary observables: two-point correlation functions, which provide a direct estimate of $\xi_D$, and the variational energy error, which provides an independent determination of the scaling exponent $\kappa$.

The most direct probe is the two-point correlation function. For a local field $\mathcal{O}$ of scaling dimension $\Delta_\mathcal{O}$ in an open chain, with one operator fixed at the center and the other at distance $r$, the finite-depth correlator has the scaling form
\begin{align}
  &
  \expval{\mathcal{O}_{L/2}\mathcal{O}_{L/2+r}}{\psi_\mathrm{var}}
  =
  C(r;L,\xi_D)
  \nonumber\\
  &
  \xrightarrow{L\gg \xi_D}
  d_\mathrm{chord}(r,L)^{-2\Delta_\mathcal{O}}
  f(r/\xi_D),
  \label{eq:corr-cft}
\end{align}
where
\begin{align}
  d_\mathrm{chord}(r,L)
  =
  \frac{L}{\pi}
  \sin(\frac{\pi r}{L}).
  \label{eq:chord}
\end{align}
For separations $r\gg\xi_D$, the scaling function develops an exponential envelope, $f(r/\xi_D)\sim\exp(-r/\xi_D)$. After factoring out the critical power-law contribution, $d_\mathrm{chord}(r,L)^{-2\Delta_\mathcal{O}}$, fitting this exponential decay directly yields the depth-dependent correlation length. Assuming the algebraic scaling relation in \cref{eq:xi-scaling}, the resulting fit yields both the non-universal amplitude $A_\xi$ and the scaling exponent $\kappa$.

The correlation function provides a direct estimate of the emergent correlation length $\xi_D$. However, within the RG framework, the same infrared scale should govern the scaling of all long-distance observables. It is therefore important to verify the extracted $\xi_D$ using an independent observable. For variational states, the energy error with respect to the target Hamiltonian provides a particularly useful second probe. We define
\begin{align}
  \delta E_\mathrm{abs}(L,D)
  =
  \expval{H_T}{\psi_\mathrm{var}}
  -
  E_0(L),
  \label{eq:absolute-energy-error}
\end{align}
where $E_0(L)$ is the exact ground-state energy of the target Hamiltonian. Because the optimized variational state is stationary within the variational manifold, the leading energy correction associated with the effective perturbation is second order in the perturbation strength $g_D$, as discussed in Appendix~\ref{app:scaling-CFT}. Conformal perturbation theory therefore predicts that, in the finite-depth regime,
\begin{equation}
  L\,\delta E_\mathrm{abs}(L,D)
  \propto
  g(L,\xi_D)^2,
  \label{eq:energy-correction-2}
\end{equation}
where $g(L,\xi_D)$ is the effective perturbation strength after coarse-graining to the scale $\xi_D$. More explicitly, conformal perturbation theory gives~\cite{cardy_conformal_1984,cardyOperatorContentTwodimensional1986,liu2024,schneiderSelfcongruentPointCritical2025}
\begin{align}
  g^2(L/\xi_D)
  \propto
  \begin{cases}
    \left(L/\xi_D\right)^2,
    & \Delta_\Phi \neq 1,
    \\
    \left(L/\xi_D\right)^2 \ln(\xi_D/a_E),
    & \Delta_\Phi = 1,
  \end{cases}
  \label{eq:running-coupling}
\end{align}
where $a_E$ is a non-universal ultraviolet cutoff. Assuming $\xi_D=A_\xi D^\kappa$, this prediction turns the energy error into a second, independent fit of $\kappa$. In the applications below, the leading perturbation has $\Delta_\Phi=1$, and the corresponding logarithmic correction is therefore included explicitly in the fits.

The scaling form in \cref{eq:running-coupling} is valid only when the finite variational correlation length is the dominant infrared cutoff, namely in the finite-depth regime $L\gg\xi_D$. When instead $\xi_D\gtrsim L$, the variational state already reproduces the finite-size critical behavior of the target Hamiltonian, so these asymptotic expressions cannot be used to extract $\kappa$. Since $\xi_D$ is precisely the unknown quantity we wish to determine, we require an independent criterion to distinguish between the finite-depth and finite-size regimes.

For this purpose, we use the bulk energy density shift,
\begin{align}
  \delta e_\mathrm{bulk}(L,D)
  =
  \frac{E(L,D)}{L}
  -
  e_0
  -
  \frac{e_\mathrm{boundary}}{L}.
  \label{eq:bulk-energy-shift}
\end{align}
For an exact critical open chain, the bulk energy density approaches its thermodynamic limit according to~\cite{affleck_universal_1986,liu2024}
\begin{align}
  \frac{E_0(L)}{L}
  =
  e_0
  +
  \frac{e_\mathrm{boundary}}{L}
  -
  \frac{\pi c v}{24L^2}
  +
  \order{L^{-3}},
  \label{eq:finite-size-energy}
\end{align}
where $c$ is the central charge and $v$ is the velocity. In the presence of a finite variational correlation length, conformal perturbation theory predicts
\begin{align}
  L^2\delta e_\mathrm{bulk}
  \simeq
  B
  \left(\frac{L}{\xi_D}\right)^2
  \ln(\frac{\min(L,\xi_D)}{a_E})
  -
  \frac{\pi c v}{24},
  \label{eq:bulk-energy-filter}
\end{align}
where $B$ is a non-universal constant; see also Appendix~\ref{app:energy-shift} for further details.

This expression provides a practical criterion for identifying the finite-depth regime. When $L^2\abs{\delta e_\mathrm{bulk}}$ remains close to the universal CFT value $\pi cv/24$, the variational state is still dominated by finite-size effects and is therefore excluded from the energy scaling analysis. Only once the bulk energy deviates appreciably from this value does the finite variational correlation length become the dominant infrared cutoff. This criterion depends only on the measured energy density and known CFT data, and not on the unknown exponent $\kappa$. In the numerical results of Section~\ref{sec:application}, we therefore use the bulk energy density shift as a practical filter to select the data included in the energy fit used to extract $\kappa$.

\section{Parameterized quantum evolutions as variational ansätze}\label{sec:parameterized}

Let us now introduce the class of parameterized quantum evolutions illustrated in~\cref{fig:partition-circuit-RG}(c), which we use as variational ansätze. Inspired by hybrid digital--analog quantum simulation approaches~\cite{daley2022practical}, we consider layered Hamiltonian evolutions in which each layer is a unitary $U(\bm{\theta})=\exp[-i H(\bm{\theta})]$ whose couplings $\bm{\theta}$ serve as variational parameters. Acting on an easily preparable initial state $\rho_0$, the circuit prepares a trial state whose parameters are optimized to minimize the energy of a target many-body Hamiltonian $H_T$,
\begin{align}
\ell(\bm{\theta},\rho_0,H_T)
=
\tr[
U(\bm{\theta})\rho_0U^\dagger(\bm{\theta})H_T
],
\label{eq:loss-function}
\end{align}
using a classical optimizer, as in variational quantum eigensolvers (VQE)~\cite{peruzzo2014,cerezo2021,endo2021,bharti2022,tilly2022}. In this work, we focus on characterizing the expressiveness of this variational class at criticality, rather than evaluating VQE performance as a ground-state search strategy.

All ansätze considered share the same layered structure. Throughout this work we consider pure-state circuits, so the initial state is $\rho_0=\ketbra{\psi_0}$, where $\ket{\psi_0}$ is a homogeneous product state. The variational state is obtained by applying a sequence of $D$ unitaries,
\begin{align}
  \ket{\psi(\bm\theta)}
  &=
  U_\mathrm{C}(\bm\theta)\,\ket{\psi_0},
  \nonumber \\
  U_\mathrm{C}(\bm\theta)
  &=
  U^{(D)}(\bm\theta_D)\cdots U^{(1)}(\bm\theta_1),
  \label{eq:circuit-skeleton}
\end{align}
where $D$ denotes the circuit depth, which plays the role of the refinement parameter introduced in Section~\ref{sec:framework}. Each layer is parameterized by a small set of Hermitian generators $\{G_k\}_{k=1}^{K}$, typically including a transverse field $G_z$ and interaction terms adapted to the target Hamiltonian. Since each layer contains only a fixed number of variational parameters, the total number of parameters scales intensively as $\order{D}$.

We consider two constructions for the layer unitary. The first is the \emph{separable} construction, corresponding to the Hamiltonian Variational Ansatz (HVA)~\cite{wecker2015,Reiner2019,Mele2022,Wiersema2020,Park2024hamiltonian} commonly used in variational quantum algorithms~\cite{cerezo2021}. Each layer applies the evolution generated by a single Hermitian operator,
\begin{align}
U_{\mathrm{sep}}(\bm{\theta})
=
\prod_{k=1}^{K}
\exp(-i\theta_k G_k),
\label{eq:Usep}
\end{align}
in a fixed order, reminiscent of QAOA~\cite{farhi_quantum_2014} and brick-wall circuits~\cite{bravo-prieto2020,jobst2022}. In contrast, the \emph{combined} construction collects non-commuting Hermitian generators inside a single exponential,
\begin{align}
U_{\mathrm{com}}(\bm{\theta})
=
\exp[-i H(\bm{\theta})],
\qquad
H(\bm{\theta})
=
\sum_{k=1}^{K}\theta_k G_k,
\label{eq:Ucom}
\end{align}
which is often a more natural description of analog and hybrid digital--analog quantum evolutions, and can provide greater variational flexibility whenever $[G_k,G_{k'}]\neq0$.

\section{Application of the framework to the transverse-field Ising model}\label{sec:application}
In this section, we apply the universal scaling
framework described above to a concrete and paradigmatic setting: the one-dimensional transverse-field Ising model (TFIM) at criticality. The critical TFIM combines a well-understood critical point with a controlled theoretical description, allowing for a precise characterization of scaling behavior. We first introduce the target Hamiltonian and its relevant properties in Section~\ref{subsec:TFIM}, then specify the variational ansätze considered in Section~\ref{subsec:ansatze}, and finally present the numerical results of the scaling analysis in Section~\ref{subsec:numerical}.

\subsection{Target Hamiltonian: TFIM~\label{subsec:TFIM}}

The target Hamiltonian is the one-dimensional TFIM at its critical point $J/h=1$, with open boundary conditions,
\begin{align}
  H_\mathrm{TFIM}
  =
  -J\sum_{n=1}^{L-1}\sigma_n^x\sigma_{n+1}^x
  -h\sum_{n=1}^{L}\sigma_n^z \,.
  \label{eq:TFIM}
\end{align}
At criticality, the low-energy physics of the model is described by the Ising conformal field theory with central charge $c=1/2$~\cite{cardy1996}. The exact ground-state energy is known analytically, since the Ising model is integrable~\cite{pfeuty_ising_1970,burkhardt1985},
\begin{align}
  E_\mathrm{TFIM}^\mathrm{OBC}(L)
  =
  1-\frac{1}{\cos(\frac{\pi}{2(2L+1)})}\, .
  \label{eq:exact_E}
\end{align}

Beyond the identity operator, the Ising universality class contains two relevant primary operators: the order parameter $\sigma(x_n)\simeq\sigma_n^x$, with scaling dimension $\Delta_\sigma=1/8$, and the energy operator $\epsilon(x_n)\simeq\sigma_n^z$, with scaling dimension $\Delta_\epsilon=1$. They are distinguished by the global $\mathbb{Z}_2$ spin-flip symmetry generated by
\begin{align}
  P=\prod_{n=1}^{L}\sigma_n^z,
\end{align}
under which $\sigma$ is $\mathbb{Z}_2$-odd, whereas $\epsilon$ is $\mathbb{Z}_2$-even.

The variational constructions used here respect this symmetry throughout. Both layer generators, the Ising coupling $\sigma_n^x\sigma_{n+1}^x$ and the transverse field $\sigma_n^z$, commute with $P$, and the initial state $\ket{\psi_0}=\bigotimes_{n=1}^{L}\ket{\uparrow}_n$ is the $\mathbb{Z}_2$-even fully polarized reference. The optimized state therefore remains in the symmetric sector, where $\expval{\sigma_n^x}=0$ identically and the order parameter cannot serve as a one-point probe of the finite-resource perturbation.

This symmetry has an important consequence for the RG description of the finite-depth state. Since the order parameter is $\mathbb{Z}_2$-odd, it cannot appear as the leading effective perturbation induced by the finite variational resources. Among the two relevant operators introduced above, only the $\mathbb{Z}_2$-even energy operator is allowed as the leading perturbation. We therefore identify $\Phi=\epsilon$, with $\Delta_\Phi=\Delta_\epsilon=1$. This fixes the correlation-length exponent to $\nu^{-1}=d-\Delta_\epsilon=1$ and places the energy scaling in the logarithmically corrected case discussed in Section~\ref{subsec:extraction}. Although the order parameter itself vanishes identically, its two-point function $\expval{\sigma_m^x\sigma_n^x}$ remains nonzero in the symmetric sector. Throughout this work we therefore use this correlator to extract the emergent correlation length $\xi_D$.

\subsection{Variational ansätze for the TFIM\label{subsec:ansatze}}

All parameterized quantum evolutions considered here share the circuit structure introduced in Section~\ref{sec:parameterized}: an ordered product of $D$ layers $U^{(d)}$ acting on the symmetric reference state $\ket{\psi_0}=\bigotimes_n\ket{\uparrow}_n$. They span two design axes. The first is the alphabet of Hermitian generators, which determines the interaction structure and connectivity. The second is the layer structure, which determines whether the generators are applied as separable Hamiltonian Variational Ansatz (HVA) sub-layers or combined inside a single exponential. Below we specify the layer unitary $U^{(d)}$ for each ansatz; the full circuit follows from Eq.~\eqref{eq:circuit-skeleton}.

\paragraph{Nearest-neighbor TFIM generators.}

The most natural choice of generators is given by the two terms of the target Hamiltonian itself, namely the Ising interaction and the transverse field,
\begin{align}
  G_{xx}
  =
  \sum_{n=1}^{L-1}\sigma_n^x\sigma_{n+1}^x,
  \qquad
  G_z
  =
  \sum_{n=1}^{L}\sigma_n^z .
  \label{eq:gen-NN}
\end{align}
Interleaving these generators as separable sub-layers gives the HVA layer for the TFIM,
\begin{align}
  U^{(d)}
  =
  \exp(-\i\theta_d^z G_z)\,
  \exp(-\i\theta_d^{xx} G_{xx}),
  \label{eq:ansatz-NN-sep}
\end{align}
whereas folding both generators into a single exponential defines the combined TFIM layer,
\begin{align}
  U^{(d)}
  =
  \exp[
  -\i\left(
  \theta_d^{xx}G_{xx}
  +
  \theta_d^zG_z
  \right)
  ].
  \label{eq:ansatz-NN-com}
\end{align}
Both ansätze contain two variational parameters per layer.

\paragraph{Kitaev generators.}

Under the Jordan--Wigner transformation, the transverse field $G_z$ becomes a fermionic chemical potential, while the Ising interaction $G_{xx}$ decomposes into equal-weight fermionic hopping and pairing terms. This naturally suggests promoting the hopping and pairing contributions to independent variational generators~\cite{kitaev_unpaired_2001},
\begin{align}
  G_\mathrm{hop}
  &=
  \sum_{n}
  \left(
  c^\dagger_n c_{n+1}
  +\mathrm{h.c.}
  \right),
  \nonumber\\
  G_\mathrm{pair}
  &=
  \sum_{n}
  \left(
  c^\dagger_n c^\dagger_{n+1}
  +\mathrm{h.c.}
  \right),
  \nonumber\\
  G_z
  &=
  \sum_{n}
  \left(
  1-2c^\dagger_n c_n
  \right).
  \label{eq:gen-Kitaev}
\end{align}
This generator set remains within the quadratic fermionic structure of the Ising problem while giving the optimizer three independent directions per layer. Using the separable HVA construction, the layer unitary becomes
\begin{align}
  U^{(d)}
  =
  \exp(-\i\theta_d^z G_z)\,
  \exp(-\i\theta_d^\mathrm{p} G_\mathrm{pair})\,
  \exp(-\i\theta_d^\mathrm{h} G_\mathrm{hop}),
  \label{eq:ansatz-Kitaev-HVA}
\end{align}
whereas the combined construction gives
\begin{align}
  U^{(d)}
  =
  \exp[
  -\i\left(
  \theta_d^\mathrm{h}G_\mathrm{hop}
  +
  \theta_d^\mathrm{p}G_\mathrm{pair}
  +
  \theta_d^zG_z
  \right)
  ].
  \label{eq:ansatz-Kitaev-com}
\end{align}
Both ansätze contain three variational parameters per layer.

\paragraph{Long-range generators.}

Finally, we extend the fermionic generators beyond nearest neighbors by dressing the hopping and pairing terms with a translationally invariant coupling kernel $f_\eta(r)$, controlled by a range parameter $\eta$. We consider two families of kernels,
\begin{align}
  f_\lambda(r)
  &=
  \lambda^{\,r-1},
  &&(\mathrm{EXP}),
  \\
  f_\alpha(r)
  &=
  r^{-\alpha},
  &&(\mathrm{POW}),
  \label{eq:kernels}
\end{align}
which define the long-range interaction generator
\begin{align}
  G_{xx}^{(\eta)}
  =
  \sum_{n<m}
  f_\eta(m-n)\,
  g_{nm}^{xx}.
  \label{eq:gxx}
\end{align}
Here $g_{nm}^{xx}$ denotes the hopping and pairing terms between sites $n$ and $m$, and the explicit Bogoliubov--de Gennes representation is given in Appendix~\ref{appsubsec:free-gaussian}. Since the Jordan--Wigner transformation is non-local, these long-range fermionic generators are not equivalent to long-range spin interactions. They therefore define a distinct variational class, naturally connected to fermionic analog and hybrid digital--analog platforms capable of implementing long-range quadratic interactions~\cite{arguello-luengo_analogue_2019,arguello-luengo_quantum_2020,arguello-luengo_engineering_2021,Arguello-Luengo2022,weiUniversalEfficientHybrid2026}.

Using the separable HVA construction, the field generator is applied in a separate sub-layer,
\begin{align}
  U^{(d)}
  =
  \exp(-\i\theta_d^z G_z)\,
  \exp(
  -\i\theta_d^{xx}
  G_{xx}^{(\eta_d)}
  ),
  \label{eq:ansatz-LR-sep}
\end{align}
whereas the combined construction folds the field generator into the same exponential,
\begin{align}
  U^{(d)}
  =
  \exp[
  -i\left(
  \theta_d^{xx}G_{xx}^{(\eta_d)}
  +
  \theta_d^zG_z
  \right)
  ],
  \qquad
  \eta\in\{\lambda,\alpha\}.
  \label{eq:ansatz-LR-com}
\end{align}
In both cases, the kernel range $\eta_d$ is itself optimized independently at every layer. Consequently, both constructions contain three variational parameters per layer. The gradient with respect to $\eta_d$ is evaluated exactly using the Fréchet derivative, as described in Appendix~\ref{app:optim-details}.
\subsection{Numerical results~\label{subsec:numerical}}

Both the target model and the Hermitian generators used to construct the parameterized evolutions are quadratic after the Jordan--Wigner transformation and preserve the $\mathbb{Z}_2$ symmetry. We therefore formulate the optimization in the Bogoliubov--de Gennes representation, where a fermionic Gaussian state on a chain of length $L$ is fully specified by a $2L\times 2L$ Dirac correlation matrix $\Gamma$~\cite{suraceFermionicGaussianStates2022}. Within this representation, state evolution, observables, gradients, and the Fubini--Study metric entering the variational principle are evaluated numerically exactly; see Appendix~\ref{app:optim-details} for details.

\subsubsection{Representative analysis: TFIM HVA\label{subsec:scalingHVA}}

We now illustrate the extraction procedure in detail for the separable TFIM HVA~\eqref{eq:ansatz-NN-sep} targeting the critical TFIM~\eqref{eq:TFIM}.

\begin{figure}[tb]
  \centering
   \includegraphics[width=0.99\linewidth]{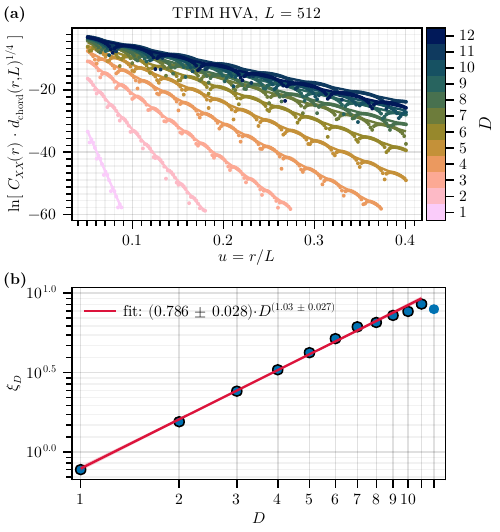}
\caption{\label{fig:TFI_corr_length}%
\textbf{Correlation-length extraction for the TFIM HVA.}
(a) Chord-corrected spin--spin correlation function for a chain of length $L=512$ at different circuit depths $D$ (color-coded). The solid lines are exponential fits to the extracted correlation envelope, from which the finite-depth correlation length $\xi_D$ is obtained.
(b) Extracted correlation length as a function of depth. Over the scaling window used in the fit, the data follow the algebraic scaling form $\xi_D=A_\xi D^\kappa$, with the fitted parameters reported in the legend.
}
\end{figure}

\paragraph{Scaling exponent from correlation functions.}

Following the procedure introduced in Section~\ref{subsec:extraction}, we first determine the finite-depth correlation length $\xi_D$ from the decay of the spin--spin correlation function. For the Ising order field, $\Delta_\sigma=1/8$, and the critical power law is removed by multiplying the lattice correlator by the appropriate chord distance to the power $2\Delta_\sigma$. With one operator fixed at the chain center and the second at distance $r$, we fit the leading finite-depth envelope,
\begin{align}
  C_{XX}(r)\,
  d_{\mathrm{chord}}(r,L)^{1/4}
  \propto
  \exp(-r/\xi_D),
  \label{eq:correlator-envelope-fit}
\end{align}
where $d_{\mathrm{chord}}(r,L)$ is defined in Eq.~\eqref{eq:chord}. The fit is performed over the bulk window $r/L\in[5\%,40\%]$, where boundary effects remain small while the finite-depth decay is clearly resolved. The exponential form in Eq.~\eqref{eq:correlator-envelope-fit} should be understood as the leading envelope used to define an operational finite-depth correlation length; subleading effects coming from both the correlation length and the lattice-scale are absorbed into the fit uncertainty.

\Cref{fig:TFI_corr_length}(a) shows the chord-corrected correlation function for the TFIM HVA~\eqref{eq:ansatz-NN-sep}. Besides the expected exponential decay, the data exhibit oscillations that are not captured by the asymptotic CFT scaling form. We attribute these oscillations to the non-equilibrium structure of the optimized finite-depth unitary state, whose correlation matrix retains features not present in the ground state of a local equilibrium Hamiltonian. To isolate the long-distance envelope, we group nearby data points in $r$ and retain the maximum within each bin. The resulting binned maxima trace the smooth upper envelope, while the troughs of the oscillations are not used to define the leading correlation length. We then extract $\xi_D$ from a linear fit to the logarithm of the envelope.

The extracted correlation lengths are shown in \cref{fig:TFI_corr_length}(b). Over the accessible scaling window they are well described by the algebraic form
\begin{align}
  \xi_D=A_\xi D^\kappa.
\end{align}
The deepest circuit is excluded from the fit because it falls outside the observed scaling trend. Possible causes include finite-size contamination, reduced reliability of the envelope extraction as $\xi_D$ approaches the available separations, or convergence to a suboptimal local minimum. The resulting exponent,
\begin{align}
    \kappa_C = 1.03 \pm 0.03,
\end{align}
provides the first estimate of the finite-depth scaling exponent. The quoted uncertainty is a jackknife standard error over the fitted depths, as detailed in Appendix~\ref{par:jackknife}. As we show below, an independent analysis based on the variational energy error yields a consistent value, supporting the central assumption of the framework that both observables are governed by the same emergent finite-depth correlation length.

\begin{figure}[tb]
  \centering
   \includegraphics[width=0.92\linewidth]{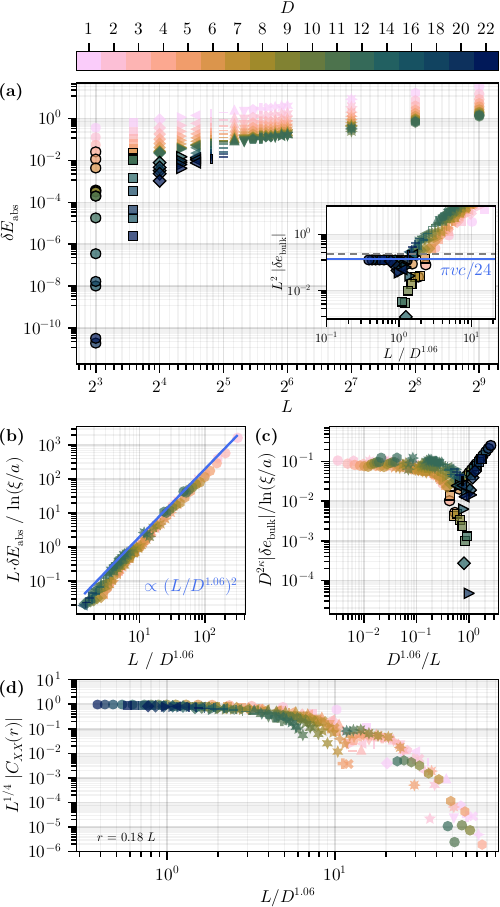}
  \caption{\label{fig:TFI_HVA-vert}%
\textbf{Extraction of the scaling exponent from the variational energy for the TFIM HVA.}
(a) Absolute energy error $\delta E_\mathrm{abs}(L,D)$ as a function of system size for different circuit depths $D$. Black-edged markers indicate calculations classified as finite-size dominated and excluded from the scaling analysis. Inset: rescaled bulk energy density $L^2\delta e_\mathrm{bulk}$ used to identify the finite-depth regime. The blue line marks the universal CFT value $\pi cv/24$, while the dashed gray line denotes the tolerance used for filtering.
(b) Energy data after division by the fitted logarithmic correction, showing the predicted quadratic dependence on the scaling variable $L/D^{\kappa_E}$, with $\kappa_E=1.06$ being the fitted value.
(c) Bulk energy density shift in the finite-depth regime, exhibiting the expected collapse after rescaling.
(d) Correlator scaling function at fixed $r/L=0.18$, demonstrating consistency with the same scaling variable extracted from the energy analysis.
}
\end{figure}

\paragraph{Scaling exponent from energies.}

The energy error provides an independent estimate of the scaling exponent $\kappa$. We fit the absolute variational energy error,
\begin{align}
  \delta E_\mathrm{abs}(L,D)
  =
  \expval{H_T}{\psi(L,D)}
  -E_0(L),
\end{align}
using the finite-depth scaling form derived in Section~\ref{subsec:extraction}. For the Ising energy perturbation, $\Delta_\Phi=1$, the predicted scaling is
\begin{align}
  L\,\delta E_\mathrm{abs}(L,D)
  =
  P\,L^2D^{-2\kappa}
  \left(\kappa\ln D+Q\right),
  \label{eq:energy-fit-model-main}
\end{align}
where $P$ is a non-universal amplitude and $Q=\ln(A_\xi/a_E)$ contains the ultraviolet cutoff entering the logarithmic correction. Since this expression is linear in $P$ for fixed $(\kappa,Q)$ we determine the optimal exponent using the VARiable PROjection (VARPRO) procedure described in Appendix~\ref{par:varpro}. The cutoff-dependent constant is constrained to its physical range using the independently extracted correlation-length amplitude $A_\xi$.

Figure~\ref{fig:TFI_HVA-vert}(a) shows the absolute variational energy error as a function of system size for different circuit depths. Increasing the depth systematically suppresses the energy error, indicating that the variational state approaches the critical ground state as additional variational resources become available. To extract the scaling exponent, however, only calculations in the finite-depth regime should be included. Following the criterion introduced in Section~\ref{subsec:extraction}, we identify calculations that remain finite-size dominated using the bulk energy density shift,
\begin{align}
  \delta e_\mathrm{bulk}
  =
  \frac{E(L,D)}{L}
  -e_0
  -\frac{e_\mathrm{boundary}}{L},
\end{align}
where $e_0=-4/\pi$ and $e_\mathrm{boundary}=1-2/\pi$ are obtained from the large-$L$ expansion of Eq.~\eqref{eq:exact_E}. For the critical TFIM,
\begin{align}
  L^2\delta e_\mathrm{bulk}
  =
  -\frac{\pi cv}{24},
\end{align}
with $c=1/2$ and $v=2$. Points whose rescaled bulk energy remains within a fixed tolerance of this universal value are classified as finite-size dominated, indicated by black-edged markers in Fig.~\ref{fig:TFI_HVA-vert}, and excluded from the fit. The inset of Fig.~\ref{fig:TFI_HVA-vert}(a) illustrates this filtering criterion. Its horizontal axis is shown as $L/D^\kappa$ for visualization only; the classification itself depends solely on $L^2\delta e_\mathrm{bulk}$ and therefore does not require prior knowledge of $\kappa$.

After excluding the finite-size dominated calculations and optimizing the scaling exponent with the VARPRO procedure, Fig.~\ref{fig:TFI_HVA-vert}(b) shows the resulting collapse of the energy data. After dividing out the fitted logarithmic correction, the remaining dependence is well described by the predicted quadratic scaling in the variable $L/\xi_D\propto L/D^\kappa$ throughout the finite-depth scaling window. Deviations become visible only as $L/\xi_D$ approaches unity, where neither the asymptotic finite-depth expansion nor leading-order conformal perturbation theory are expected to remain quantitatively accurate.

Panels (c) and (d) provide complementary consistency checks of the extracted correlation length. Panel (c) shows the corresponding collapse of the bulk energy density shift in the finite-depth regime. In this representation, the rescaled data approach a non-universal plateau corresponding to the coefficient with which $\xi_D^{-2}$ contributes to the bulk energy density, in analogy with the universal finite-size coefficient $\pi cv/24$ multiplying $L^{-2}$ in the exact critical chain. Panel (d) demonstrates that the spin--spin correlation function collapses when expressed in terms of the same scaling variable $L/D^\kappa$. The resulting exponent,
\begin{align}
    \kappa_E = 1.06 \pm 0.07,
\end{align}
is consistent with the independent estimate obtained from the correlation-length scaling, supporting the central assumption of the framework that both observables are governed by the same emergent finite-depth correlation length.

\begin{figure}[tb]
  \centering
   \includegraphics[width=\linewidth]{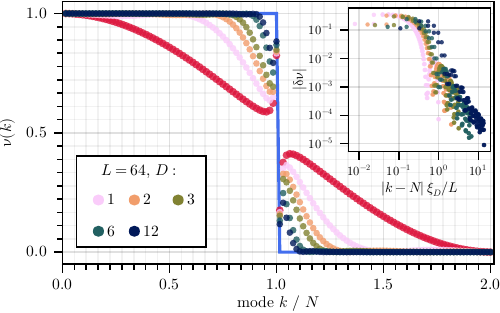}
  \caption{\label{fig:BZ-TFI_comb}
\textbf{Quasiparticle occupation analysis for the TFIM HVA.}
Occupation numbers $\nu_k(\Gamma_{\mathrm{var}})$ of the variational state in the eigenbasis of the target Hamiltonian, plotted as a function of the normalized mode index $k/N$ for a chain of length $L=64$. The solid blue curve denotes the exact ground-state occupation, while the red curve corresponds to the initial product state $\ket{\uparrow}^{\otimes N}$. Increasing the circuit depth progressively reconstructs the Fermi step, leaving the remaining error localized around the gap-closing point. Inset: occupation error $\abs{\delta\nu}=\abs{\nu_k(\Gamma_{\mathrm{var}})-\nu_k(\Gamma_{\mathrm{GS}})}$ plotted as a function of the rescaled distance from the Fermi surface, $\abs{k-N}\xi_D/L$, where $\xi_D$ is extracted from the correlation-function analysis of Fig.~\ref{fig:TFI_corr_length}(b). The collapse demonstrates that the width of the unresolved low-energy region is controlled by the same finite-depth correlation length.
}
\end{figure}

\paragraph{Quasiparticle analysis.}

Finally, we examine the quasiparticle occupations, which provide a microscopic view of how finite variational resources limit the approximation of the critical ground state. We evaluate the variational correlation matrix in the eigenbasis $V_t$ of the target Hamiltonian,
\begin{align}
  \nu_k(\Gamma)
  =
  \left(\mathrm{diag}\,V_t^\dagger\Gamma V_t\right)_k,
  \label{eq:mode-occupation}
\end{align}
where the modes are ordered by increasing single-particle energy. The exact ground state corresponds to a sharp Fermi step, $\nu_k(\Gamma_{\mathrm{GS}})=\Theta(N-k)$, with the first $N$ modes occupied.

As shown in Fig.~\ref{fig:BZ-TFI_comb}, increasing the circuit depth progressively reconstructs this step everywhere except within a narrowing window around the gap-closing point. When the occupation error $\abs{\nu_k-\Theta(N-k)}$ is plotted as a function of the rescaled distance $\abs{k-N}\xi_D/L$, the data collapse onto a single curve. This demonstrates that the width of the unresolved low-energy region is governed by the same finite-depth correlation length extracted independently from both the correlation functions and the energy scaling.

This quasiparticle analysis provides a microscopic interpretation of the RG picture developed in Section~\ref{sec:framework}. Finite variational resources leave modes far from the gap closing essentially resolved already at shallow depth, while the remaining error becomes increasingly localized around the modes responsible for the long-distance critical correlations. As the circuit depth increases, this unresolved momentum window shrinks proportionally to $\xi_D^{-1}$, explaining why the same emergent correlation length controls both the real-space correlations and the energy scaling. Although this analysis does not provide an independent estimate of $\kappa$, it makes the finite-depth infrared cutoff visible at the level of the quasiparticle occupations.

\subsubsection{Comparison across variational ansätze}\label{subsec:comparison}

Having established the scaling analysis in detail for the TFIM HVA, we now apply the same protocol to the remaining variational ansätze introduced in Section~\ref{subsec:ansatze}. The purpose of the comparison is to demonstrate how the above detailed scaling framework quantitatively distinguishes different variational architectures through a common observable, the scaling exponent $\kappa$. Since the numerical procedures closely parallel those discussed above, the corresponding correlation and energy scaling plots are presented in Appendix~\ref{app:detailed-plots}. Here we focus on the resulting scaling exponents.

\begin{table*}[th]
  \setlength{\tabcolsep}{8pt}
  \begin{tabular}{l l|l|l|r|l}
    \textbf{Ansatz} & & $\bm{\kappa_C}$ & $\bm{\kappa_E}$ & $n_D^{\geq 3}$ & $D$ range $(n_D^{\geq 3})$ \\
    \hline
    TFIM HVA        & \cref{eq:ansatz-NN-sep}                & $1.03 \pm 0.027$ & $1.06 \pm 0.071$ & $14$ & 1 to 18 \\[0.25ex]
    TFIM combined   & \cref{eq:ansatz-NN-com}                & $1.19 \pm 0.013$ & $1.41 \pm 0.025$ & $11$ & 2 to 12 \\[0.25ex]
    Kitaev HVA      & \cref{eq:ansatz-Kitaev-HVA}            & $1.04 \pm 0.14$  & $1.23 \pm 0.10$  & $11$ & 2 to 12 \\[0.25ex]
    Kitaev combined & \cref{eq:ansatz-Kitaev-com}            & $1.48 \pm 0.35$  & $1.21 \pm 0.13$  & $10$ & 3 to 12 \\[0.25ex]
    EXP HVA ($\eta=\lambda$)      & \cref{eq:ansatz-LR-sep} & $2.08 \pm 0.074$ & $2.62 \pm 0.18$  & $5$  & 1 to 5  \\[0.25ex]
    EXP combined ($\eta=\lambda$) & \cref{eq:ansatz-LR-com} & $3.07 \pm 0.13$  & $3.26 \pm 0.29$  & $4$  & 2 to 5  \\[0.25ex]
    POW HVA ($\eta=\alpha$)       & \cref{eq:ansatz-LR-sep} & $1.05 \pm 0.041$ & $1.02 \pm 0.065$ & $11$ & 2 to 12 \\[0.25ex]
    POW combined ($\eta=\alpha$)  & \cref{eq:ansatz-LR-com} & $1.56 \pm 0.21$  & $1.65 \pm 0.14$  & $11$ & 2 to 12
  \end{tabular}
  \caption{\label{tab:kappas}%
\textbf{Scaling exponents extracted for the variational ansätze considered}. The exponent $\kappa_C$ is obtained from the scaling of the correlation length, while $\kappa_E$ is obtained from the finite-depth scaling of the energy error. Quoted uncertainties are jackknife standard errors over the fitted circuit depths. The column $n_D^{\geq3}$ reports the number of circuit depths for which at least three system sizes satisfy the finite-depth criterion and therefore contribute to the energy fit. The final column gives the corresponding depth range.}
\end{table*}

Table~\ref{tab:kappas} reports two independent exponent estimates for each ansatz: $\kappa_C$, extracted from the correlation-length scaling, and $\kappa_E$, extracted from the finite-depth scaling of the energy error. The final two columns provide a simple robustness diagnostic for the energy fit. The quantity $n_D^{\geq3}$ counts the number of circuit depths for which at least three different system sizes remain in the finite-depth regime and therefore enter the fit meaningfully; the final column gives the corresponding depth range. This diagnostic is important because the energy collapse is controlled only when $L\gg\xi_D$. For rapidly improving ansätze, $\xi_D$ quickly becomes a sizable fraction of the available system sizes, leaving fewer depths from which the finite-depth scaling can be extracted.

Within the accessible scaling windows, all ansätze are compatible with algebraic growth of the induced correlation length. The two independent estimates, $\kappa_C$ and $\kappa_E$, generally agree within the limitations of the available data. This consistency supports the central assumption underlying the framework, namely that a single emergent finite-depth correlation length governs the long-distance behavior of different observables. The agreement is particularly clear for the TFIM HVA, where both methods yield essentially identical exponents. Larger discrepancies appear for the remaining ansätze, especially those for which only a limited number of circuit depths satisfy the finite-depth criterion. In these cases, the reduced scaling window limits the precision of the energy analysis rather than necessarily indicating a breakdown of the scaling framework.

The quoted uncertainties are jackknife standard errors over the set of fitted depths. They quantify the sensitivity of the fit to omitting one depth at a time, but they should not be read as the full systematic uncertainty. For the exponentially interacting ansätze in particular, the rapid growth of the correlation length leaves only a few circuit depths well inside the finite-depth regime, limiting the leverage of the energy analysis. Larger system sizes would be required to determine whether the fitted exponents have fully reached their asymptotic values. We therefore interpret the largest exponents in Table~\ref{tab:kappas} as estimates within the accessible scaling window rather than as final asymptotic numbers.

The most robust qualitative trend is the contrast between the two long-range kernel families. The exponential ansätze produce substantially larger fitted exponents, with values $\kappa\gtrsim2$ in the accessible window, whereas the power-law ansätze remain closer to the nearest-neighbor results, around $\kappa\simeq1$ for the HVA version and $\kappa\simeq3/2$ for the combined version. Thus, long-range support alone is not sufficient to guarantee a large scaling advantage: what matters is not merely connecting distant sites, but how the variational resource resolves length scales.
A plausible explanation is motivated by approximation theory~\cite{beylkinApproximationFunctionsExponential2005}: exponential interactions expose independently tunable decay scales at the level of the layer parameterization, providing a natural finite-depth dictionary of length scales.
Power-law interactions, by contrast, encode scale structure more globally in a single scale-free profile. This mechanism would account for the superior scaling observed for the exponential ansätze here, as well as related finite-size analyses in the literature~\cite{tabares2023,tabares_programming_2026}.

A useful caveat applies to the power-law cases. The energy-collapse analysis assumes that the finite-depth perturbation is governed by the $\mathbb{Z}_2$-even Ising energy operator with scaling dimension $\Delta_\epsilon=1$. For sufficiently slowly decaying power-law couplings, however, the ansatz may explore generators whose low-energy physics departs from the short-range Ising universality class~\cite{koffelEntanglementEntropyLongRange2012,vodola_kitaev_2014,schneiderEntanglementSpectrumQuantum2022,defenu2023}. While this does not affect the operational extraction of $\kappa_C$ from the correlation functions, it suggests interpreting the corresponding energy exponents with appropriate caution.

A second architectural trend concerns the construction of the circuit layers. Across most generator families, the combined ansätze achieve larger fitted exponents than their separable HVA counterparts. This trend should not be read as a strict manifold-inclusion statement. Rather, the two constructions generate different variational manifolds at the same layer count. In the combined case, the interaction and transverse-field terms compete inside a single exponential, so a layer can already encode part of the gap-closing structure relevant to the target critical point. In the separable case, the same competition is generated through the composition of sub-layers, with the corresponding commutator structure appearing at higher order. This provides a plausible explanation for the larger combined exponents observed in most kernel families.

Throughout this comparison we count resources by the number of layers $D$. Because the ansätze differ in parameter count and in whether a layer is realized as one combined evolution or several separable evolutions, comparisons at fixed parameter number, fixed total evolution time, or fixed compilation overhead~\cite{weiUniversalEfficientHybrid2026} may change non-universal prefactors. The scaling exponent $\kappa$ nevertheless provides a common diagnostic of how efficiently each variational architecture converts depth into correlation length.

Taken together, these results demonstrate that applying the universal scaling framework to parameterized quantum evolutions provides a consistent quantitative comparison of their expressiveness at criticality. Across the variational ansätze considered, the scaling exponents extracted from correlation functions and energy scaling exhibit good overall agreement within the accessible depth window, while revealing clear architectural trends. Exponential long-range kernels convert depth into infrared resolution most efficiently among the families studied, whereas power-law kernels remain much closer to nearest-neighbor behavior. Combined layers generally outperform separable HVA layers, suggesting that both the interaction profile and the organization of generators inside each variational layer play an important role in determining the finite-depth scaling exponent.

\section{Conclusions and outlook}\label{sec:conclusion}

We have adapted universal finite-resource scaling ideas established in tensor-network theory to quantify the efficiency of finite-depth parameterized quantum evolutions as variational ansätze for critical ground states. The central object is the ansatz-induced correlation length $\xi_D$, whose algebraic growth with the number of layers, $\xi_D\sim D^\kappa$, measures how efficiently depth is converted into long-distance quantum correlations. The framework thereby places parameterized evolutions in the same finite-resource scaling language as tensor-network approximations, while remaining agnostic to the microscopic structure of the ansatz. We applied it to the critical transverse-field Ising model, where free-fermion methods give access to large system sizes. There, correlation functions and the energy error yield independent estimates governed by the same infrared length, and all ansätze are compatible with algebraic growth of $\xi_D$ within the accessible scaling windows. The fitted exponents nevertheless differ substantially: exponential long-range kernels are the most efficient, power-law kernels remain close to nearest-neighbor behavior, and combined layers generally outperform their separable HVA counterparts. Neither long-range support nor generator range alone is therefore decisive. A complementary quasiparticle analysis locates the finite-depth error in an unresolved window around the low-energy gap-closing modes, whose width shrinks as $\xi_D^{-1}$. This realizes the RG picture microscopically: finite depth leaves the ultraviolet structure essentially resolved while acting as an infrared cutoff on the critical fixed point.

Several directions naturally follow from this work. The largest exponents, found for the exponential ansätze, should be confirmed at larger system sizes, where a wider finite-depth regime $L\gg\xi_D$ would allow a cleaner asymptotic extraction. More broadly, an analytic theory connecting the algebra of the available generators to the exponent $\kappa$ would be valuable, as would a link between $\kappa$ and geometric quantities controlling trainability, such as the rank and conditioning of the quantum Fisher information matrix and the associated dynamical Lie algebra~\cite{cerezo2021,larocca_diagnosing_2022,larocca_theory_2023,larocca_barren_2025}. Finally, the framework is not restricted to the Ising model or to free fermions. A systematic catalog of $\kappa$ across interacting and higher-dimensional critical models, tensor-network-inspired circuits, and hybrid analog--digital architectures could turn finite-resource scaling into a practical benchmark for critical-state preparation and a guide for designing hardware-native ansätze whose finite resources are spent on the infrared degrees of freedom that matter most.

\begin{acknowledgments}
We thank Alberto~Muñoz de las Heras and Diego~Porras for valuable discussions during the early stages of this work.
We acknowledge support from the Proyecto Sinérgico CAM Programa TEC-2024/COM-84 QUITEMAD-CM; the CSIC Research Platform on Quantum Technologies PTI-001; the grant PID2024-160172NB-I00 funded by MICIU/AEI/10.13039/501100011033 and by FEDER/UE; and the Spanish project PID2024-162384NB-I00 funded by MICIU/AEI/10.13039/501100011033 and by FEDER/UE.
AGT acknowledges support from the QuantERA project MOLAR, reference PCI2024-153449, funded by MICIU/AEI/10.13039/501100011033 and by the European Union, and from the Programa Fundamentos FBBVA through grant EIC24-1-17304.
JTS also acknowledges support from the Programa Fundamentos FBBVA through grant EIC24-1-17304.
CT acknowledges support from Comunidad de Madrid (PIPF-2022/TEC-25625) and from Fundación Humanismo y Ciencia.
The authors also acknowledge the support of the Centro de Supercomputación de Galicia (CESGA) providing access to the HPC facility FinisTerrae III as well as the Unidad de Sistemas Cálculo of the CSIC providing access to the HPC facility Drago for performing the numerical simulations.
\end{acknowledgments}

%%% Create the reference section using BibTeX:
\bibliography{variational_LR_paper.bib}

% \clearpage

% \onecolumn
\appendix

\section{Renormalization-group perspective on finite-resource scaling\label{app:RG}}

This appendix summarizes the renormalization-group (RG) arguments underlying the finite-resource scaling analysis used in Section~\ref{sec:framework}. The central idea is that finite variational resources act as an effective relevant perturbation of a critical fixed point, generating an emergent correlation length in direct analogy with finite-size and finite-temperature effects. We first briefly review the RG description of critical systems before deriving the corresponding scaling forms.

At a quantum critical point, the long-distance physics is invariant under changes of length scale. The RG~\cite{wilson1975,cardy1996} provides the formal framework for describing this scale invariance. In the Euclidean path-integral representation, illustrated in Fig.~\ref{fig:partition-circuit-RG}(a), the ground state is represented by a two-dimensional spacetime partition function with microscopic spacings $a$ in space and $\delta$ in imaginary time. An RG transformation coarse grains both directions by a scale factor $b$,
\begin{align}
a &\rightarrow a'=ba,\\
\delta &\rightarrow \delta'=b\delta,
\label{eq:RG-rescaling}
\end{align}
while preserving the long-distance physics. Repeated coarse graining defines a flow in the space of Hamiltonians. Critical Hamiltonians correspond to unstable fixed points of this flow, whereas generic gapped phases flow toward stable trivial fixed points.

The behavior close to a critical fixed point is determined by the scaling operators $\hat{o}$, defined through
\begin{align}
\hat{o}\rightarrow \hat{o}'=b^{-\Delta_o}\hat{o},
\label{eq:RG-scaling}
\end{align}
where $\Delta_o$ is the scaling dimension. A perturbation of the Hamiltonian by a local operator,
\begin{align}
H = H^*+g \int \hat{o} \dd{x},
\end{align}
therefore induces the RG flow
\begin{align}
g(b)\propto b^{d-\Delta_o},
\label{eq:RG-coupling-flow}
\end{align}
where $d$ is the spacetime dimension. Perturbations with $\Delta_o<d$ grow under coarse graining and are therefore \emph{relevant}, while those with $\Delta_o>d$ are \emph{irrelevant}. A relevant perturbation drives the system away from the critical fixed point and generates a finite correlation length,
\begin{align}
\xi\propto g^{-\nu},
\qquad
\nu=\frac{1}{d-\Delta_o}.
\label{eq:xi-relevant}
\end{align}

The same RG picture provides a natural interpretation of finite-resource variational approximations of critical ground states. Rather than viewing finite circuit depth as a generic approximation error, we regard it as an effective relevant perturbation that displaces the variational state away from the critical fixed point. The finite variational resource therefore induces an emergent correlation length $\xi_D$, analogous to the correlation lengths generated by finite system size or finite temperature. The central hypothesis of this work is that the scaling of $\xi_D$ with the available resources provides a universal measure of the expressiveness of a variational ansatz. We now show how this picture naturally recovers the familiar finite-size and finite-temperature scaling forms before extending them to finite-resource scaling.

\paragraph{Finite-size scaling.}
A finite system provides the simplest example of an infrared cutoff. For a lattice model with $N=L/a$ sites, the RG transformation successively reduces the number of degrees of freedom and can therefore be iterated only until the coarse-grained system contains $\order{1}$ sites. Consequently, the total rescaling factor is bounded by
\begin{align}
  b_{\max}\sim\frac{L}{a}.
  \label{eq:b-max-fss}
\end{align}
From the RG perspective, this finite cutoff prevents the flow from reaching the critical fixed point. Instead, it is equivalent to introducing an effective relevant perturbation whose strength is set by the largest accessible length scale,
\begin{align}
  g_{\mathrm{eff}}
  \propto
  b_{\max}^{-(d-\Delta_o)}
  \propto
  \left(\frac{L}{a}\right)^{-1/\nu},
  \label{eq:g-eff-fss}
\end{align}
where we used Eq.~\eqref{eq:RG-coupling-flow}. The corresponding correlation length is therefore limited by the system size,
\begin{align}
  \xi_L
  \propto
  g_{\mathrm{eff}}^{-\nu}
  \propto
  L.
  \label{eq:xi-fss}
\end{align}
The finite critical system is thus equivalent, within the scaling regime, to an infinite system perturbed away from criticality by a coupling $g_{\mathrm{eff}}\sim L^{-1/\nu}$. Universal observables consequently depend only on the ratio between the two infrared length scales, $L/\xi(g)$, leading to the familiar finite-size scaling forms discussed in the main text.

\paragraph{Finite-temperature scaling.}
Finite temperature provides a second example of an infrared cutoff. In the Euclidean path-integral representation, illustrated in Fig.~\ref{fig:partition-circuit-RG}(a), the imaginary-time direction has a finite extent $\beta$, so the RG flow is truncated once the coarse-grained temporal dimension reaches $\order{1}$. This introduces the thermal correlation length
\begin{align}
  \xi_\beta=v\,\beta^{1/z},
  \label{eq:xi-thermal}
\end{align}
where $v$ is the velocity of the low-energy excitations and $z$ the dynamical critical exponent. When $\xi_\beta<L$, thermal fluctuations dominate the infrared physics, and universal observables depend on the two dimensionless ratios $L/\xi(g)$ and $L/\xi_\beta$, leading to the standard finite-temperature scaling forms.

\paragraph{Finite-resource scaling.}
Finite variational resources provide a third infrared cutoff. The central hypothesis of this work is that a finite-depth parameterized quantum evolution can be interpreted as an effective relevant perturbation of the critical fixed point. Consequently, the finite resource induces an emergent correlation length $\xi_D$, which truncates the RG flow in exactly the same way as the finite system size or the finite temperature.

Universal observables therefore depend only on the ratios between the three competing infrared scales, $L$, $\xi_\beta$, and $\xi_D$, together with the correlation length $\xi(g)$ associated with deviations from criticality. Restricting to zero temperature leaves the two-scale problem discussed in the main text, where observables become universal functions of $L/\xi_D$. The limiting regimes $\xi_D\gg L$ and $\xi_D\ll L$, together with the scaling forms used to extract the exponent $\kappa$, are discussed in Section~\ref{subsec:extraction}.

\section{Optimization procedure\label{app:optim-details}}

This appendix summarizes the optimization procedure used throughout this work. All variational states are optimized by imaginary-time evolution projected onto the variational manifold using the quantum natural gradient. Since both the target Hamiltonian and all variational generators are quadratic fermionic operators, every quantity entering the optimization, including the energy, its gradient, and the Fubini--Study metric, can be evaluated exactly within the fermionic Gaussian formalism. We first briefly review the natural-gradient equations in Section~\ref{appsubsec:natural} before describing their efficient implementation for free Gaussian fermionic models in Section~\ref{appsubsec:free-gaussian}.

\subsection{Natural-gradient optimization\label{appsubsec:natural}}

Any variational ansatz defines a manifold of quantum states parameterized by a set of variational parameters $\bm{\theta}$,
\begin{align}
M=\{\ket{\psi(\bm{\theta})}\}\,.
\end{align}

To optimize these parameters we employ variational imaginary-time evolution based on the McLachlan variational principle~\cite{mclachlan1964,popov2024}. The exact imaginary-time evolution of the density matrix,
\begin{align}
\pdv{\rho}{\tau}
=
-\acomm{H}{\rho}
+
2\expval{H}_\rho\,\rho
\eqqcolon
\mathcal L[\rho],
\label{eq:imag-time-evo}
\end{align}
is projected onto the tangent space of the variational manifold by minimizing the distance between the exact evolution and its variational approximation.

The McLachlan variational principle determines the parameter velocities by minimizing the distance between the exact and variational imaginary-time evolution,
\begin{align}
    \norm{
    \sum_n
    \pdv{\rho}{\theta_n}
    \pdv{\theta_n}{\tau}
    -
    \mathcal{L}[\rho]
    }^2 .
\end{align}
This projection yields the variational imaginary-time evolution equation
\begin{align}
    2\sum_m
    g_{nm}
    \pdv{\theta_m}{\tau}
    =
    \Tr(
    \pdv{\rho}{\theta_n}
    \mathcal{L}[\rho]
    ),
    \label{eq:variational-Schroedinger}
\end{align}
where $g_{nm}$ is the Fubini--Study metric on the variational manifold. For pure states, the Fubini--Study metric is given by
\begin{align}
g_{nm}
&=
\Re(G_{nm}),
\label{eq:metric}\\
G_{nm}
&=
\braket{\pdv{\psi}{\theta_n}}{\pdv{\psi}{\theta_m}}
-
\braket{\pdv{\psi}{\theta_n}}{\psi} \braket{\psi}{\pdv{\psi}{\theta_m}} \nonumber \\
&=
\frac{I_{nm}}{4},
\end{align}
where $I_{nm}$ denotes the quantum Fisher information matrix. For pure states, the right-hand side of Eq.~\eqref{eq:variational-Schroedinger} reduces to the energy gradient,
\begin{align}
\Tr(
\pdv{\rho}{\theta_n}
\mathcal{L}[\rho]
)
=
- \pdv{\expval{H}}{\theta_n},
\end{align}
so that the optimization corresponds to an imaginary-time natural-gradient flow toward a stationary point of the variational energy. Rather than discretizing Eq.~\eqref{eq:variational-Schroedinger} with a fixed-step Euler scheme, we integrate it using an adaptive Dormand--Prince fourth/fifth-order Runge--Kutta method~\cite{calvo_fifth-order_1990,suli2003}, which provides improved numerical stability and convergence. In practice, this natural-gradient integration acts as a strong preconditioner: the Fubini--Study metric rescales the descent direction according to the local geometry of the manifold and carries the state rapidly through the bulk of the optimization landscape. The final approach to the local minimum is markedly slower, because the metric becomes increasingly flat in this region and the natural-gradient direction loses much of its advantage. An adaptive line search along the update direction mitigates this slowdown, but such schemes are constructed for near-Euclidean geometries. We therefore complete the last, short segment of the trajectory using a plain Euclidean gradient descent. This is justified because any smooth manifold is locally flat. Over the small remaining parameter distance the Fubini--Study metric is effectively constant, so the natural and Euclidean gradients differ only by a fixed linear rescaling and share the same stationary points. The Euclidean gradient therefore provides a valid and inexpensive descent direction for the final convergence.

\subsection{Efficient implementation for fermionic Gaussian states}
\label{appsubsec:free-gaussian}

The optimization described above becomes numerically efficient because both the target Hamiltonian and all variational generators considered in this work are quadratic fermionic operators. Consequently, every variational state remains a fermionic Gaussian state throughout the evolution and can be represented exactly by its correlation matrix. This avoids the exponential growth of the Hilbert space and enables exact evaluation of the energy, its gradients, and the Fubini--Study metric for systems containing several hundred sites. In this section we summarize the Gaussian formalism underlying our implementation, following the notation of Ref.~\cite{suraceFermionicGaussianStates2022}.

A fermionic Gaussian state on a lattice with $N$ sites is completely characterized by its two-point correlation functions. We collect them into the Dirac correlation matrix
\begin{align}
\Gamma=
\begin{pmatrix}
\expval{c_n^\dagger c_m} &
\expval{c_n^\dagger c_m^\dagger}\\
\expval{c_n c_m} &
\expval{c_n c_m^\dagger}
\end{pmatrix},
\end{align}
written in the Nambu basis. Likewise, every quadratic Hamiltonian admits the Bogoliubov--de Gennes representation
\begin{align}
  H = \sum_{n,m=1}^{N} &\big( A_{n,m} c^\dagger_n c_m - A^*_{n,m}c_n c^\dagger_m \nonumber  \\
  &+ B_{n,m} c_n c_m - B^*_{n,m}c^\dagger_n c^\dagger_m \big) \,,
\end{align}
where $A = A^\dagger$, and $B^T = -B$.
Representing both states and generators in the Bogoliubov--de Gennes form allows every layer unitary to be evaluated efficiently through the eigendecomposition of its single-particle Hamiltonian. This is particularly important for the combined ansätze introduced in Section~\ref{subsec:ansatze}, where a layer is generated by a linear combination of non-commuting operators,
\begin{align}
    h_s=\sum_k \theta_{s,k}G_{s,k},
\end{align}
so that the derivative of the corresponding unitary cannot be obtained by differentiating each exponential separately. Instead, it requires the Fréchet derivative of the matrix exponential.

Writing the eigendecomposition of the layer Hamiltonian as
\begin{align}
    h_s=V_sD_sV_s^\dagger,
\end{align}
with
\begin{align}
    D_s=\mathrm{diag}(d_1,\ldots,d_{2N}),
\end{align}
the layer propagator is
\begin{align}
    U_s=\exp(-\i h_s).
\end{align}
Denoting matrices expressed in the eigenbasis of $h_s$ by
\begin{align}
    \widetilde A=V_s^\dagger A V_s,
\end{align}
the derivative of the propagator with respect to the parameter $\theta_{s,k}$ is given exactly by Duhamel's formula,
\begin{align}
\pdv{U_s}{\theta_{s,k}}
&=
-\i
\int_0^1
e^{-\i(1-\tau)h_s}
G_{s,k}
e^{-\i\tau h_s}
\,d\tau
\nonumber\\
&=
-\i
V_s
\left(
\Phi_s^{-}
\odot
\widetilde G_{s,k}
\right)
V_s^\dagger,
\label{eq:duhamel}
\end{align}
where $\odot$ denotes element-wise multiplication and the Fréchet kernel $\Phi_s^{-}$ is the divided difference of the negative exponential,
\begin{align}
\left(\Phi_s^{-}\right)_{mn}=
\begin{cases}
\dfrac{e^{-\i d_m}-e^{-\i d_n}}
{-\i(d_m-d_n)},
&
m\neq n,
\\[1em]
e^{-\i d_m},
&
m=n.
\end{cases}
\end{align}
It is useful to introduce the Hermitian tangent generator at the output of layer \(s\),
\begin{align}
K_{s,k}
&=
\i
\pdv{U_s}{\theta_{s,k}}
U_s^\dagger
\nonumber\\
&=
V_s
\left[
\left(\Phi_s^{-}\odot\widetilde G_{s,k}\right)e^{\i D_s}
\right]
V_s^\dagger .
\label{eq:fgs-tangent-generator}
\end{align}
Thus, once the eigendecomposition of $h_s$ has been computed, both the layer propagator and all of its tangent generators are obtained exactly without numerical differentiation.

The availability of the layer propagators and their Fréchet derivatives enables an efficient implementation of the natural-gradient optimization. Rather than differentiating the full circuit repeatedly, the energy, its gradient, and the Fubini--Study metric are all obtained through a single forward/backward sweep over the variational circuit. The forward sweep propagates the correlation matrix through the layers, while the backward sweep propagates an adjoint matrix encoding the derivative of the energy with respect to the intermediate correlation matrices.

\paragraph{Forward sweep.}
Starting from the initial correlation matrix $\Gamma_0$, each variational layer acts by a Bogoliubov transformation,
\begin{align}
\Gamma_s
=
U_s\Gamma_{s-1}U_s^\dagger,
\qquad
U_s=e^{-\i h_s},
\label{eq:fgs-forward}
\end{align}
where $h_s$ is the Bogoliubov--de Gennes matrix generating layer $s$. During the forward sweep we store the intermediate correlation matrices together with the eigendecompositions of each layer Hamiltonian. These quantities are subsequently reused in the gradient and metric evaluation, avoiding repeated diagonalizations.

\paragraph{Energy evaluation.}
For a quadratic target Hamiltonian, the variational energy is linear in the correlation matrix and can therefore be written as
\begin{align}
\ell(\bm{\theta})
=
\Tr(\Lambda_0\Gamma_D),
\end{align}
where $\Lambda_0$ is determined solely by the target Hamiltonian. This linear form makes the subsequent adjoint evaluation particularly efficient.

\paragraph{Backward sweep: adjoint gradient.}
We define the adjoint cost matrix \(\Lambda_s\) by back-propagating \(\Lambda_0\) through the same layer unitaries,
\begin{align}
  \Lambda_{s-1} = U_s^{\dagger}\,\Lambda_s\,U_s.
  \label{eq:fgs-adjoint}
\end{align}

The energy gradient with respect to a parameter \(\theta_{s,k}\) of layer \(s\) is then a single trace evaluated at stage \(s\). Since
\(\partial_{\theta_{s,k}}\Gamma_s=-\i\comm{K_{s,k}}{\Gamma_s}\), it reads
\begin{align}
  \pdv{\ell}{\theta_{s,k}}
  &=
  -\i\Tr\left(
  \Lambda_s\comm{K_{s,k}}{\Gamma_s}
  \right)
  \nonumber\\
  &=
  2\,\mathfrak{Im}\,\Tr\left[
  \widetilde{\Lambda}_s
  \left(\Phi_s^{-}\odot\widetilde G_{s,k}\right)
  \widetilde{\Gamma}_{s-1}e^{\i D_s}
  \right],
  \label{eq:fgs-grad-combined}
\end{align}
where the tildes denote rotation into the eigenbasis of \(h_s\), \(\widetilde A = V_s^{\dagger} A\, V_s\). In a separable layer, \(h_s=\theta_{s,k}G_{s,k}\), so \(G_{s,k}\) commutes with \(h_s\) and \cref{eq:fgs-tangent-generator} reduces to \(K_{s,k}=G_{s,k}\). Consequently,
\begin{align}
  \pdv{\ell}{\theta_{s,k}}
  =
  -\i\Tr\left(
  \Lambda_s\comm{G_{s,k}}{\Gamma_s}
  \right),
  \label{eq:fgs-grad-separable}
\end{align}
which is the familiar commutator form for a unitary derivative. Thus, once the forward sweep has been completed, every variational parameter contributes only a single trace evaluation, yielding the complete energy gradient without finite-difference approximations.

\paragraph{Fubini--Study metric.}
The same forward sweep yields the Fubini--Study metric tensor \(g_{kl}\) of \cref{eq:metric}. Using Wick's theorem in the Nambu-doubled convention above, the entry coupling parameters \(k\) and \(l\) (with \(s_k\le s_l\)) is
\begin{align}
  g_{kl}
  =
  \frac{1}{2}\mathfrak{Re}\Tr\left[
  K_{s_k,k}\Gamma_{s_k}
  K_{s_l,l}^{(l\to s_k)}
  \left(\mathbb{I}-\Gamma_{s_k}\right)
  \right],
  \label{eq:fgs-metric}
\end{align}
where
\begin{align}
K_{s_l,l}^{(l\to s_k)}
=
U_{s_k+1}^{\dagger}\cdots U_{s_l}^{\dagger}
K_{s_l,l}
U_{s_l}\cdots U_{s_k+1}
\end{align}
is the tangent generator of parameter \(l\), including its within-layer Fréchet dressing, backward-evolved in the Heisenberg picture to the output of layer \(s_k\). The entries for \(s_l<s_k\) follow from the symmetry \(g_{kl}=g_{lk}\).

The forward and backward sweeps therefore provide the energy, its gradient, and the Fubini--Study metric simultaneously, so that each optimization step requires only a single traversal of the circuit.

\subsection{Optimized variational parameters}

\begin{figure*}[tb]
  \centering
  \includegraphics[width=0.99\textwidth]{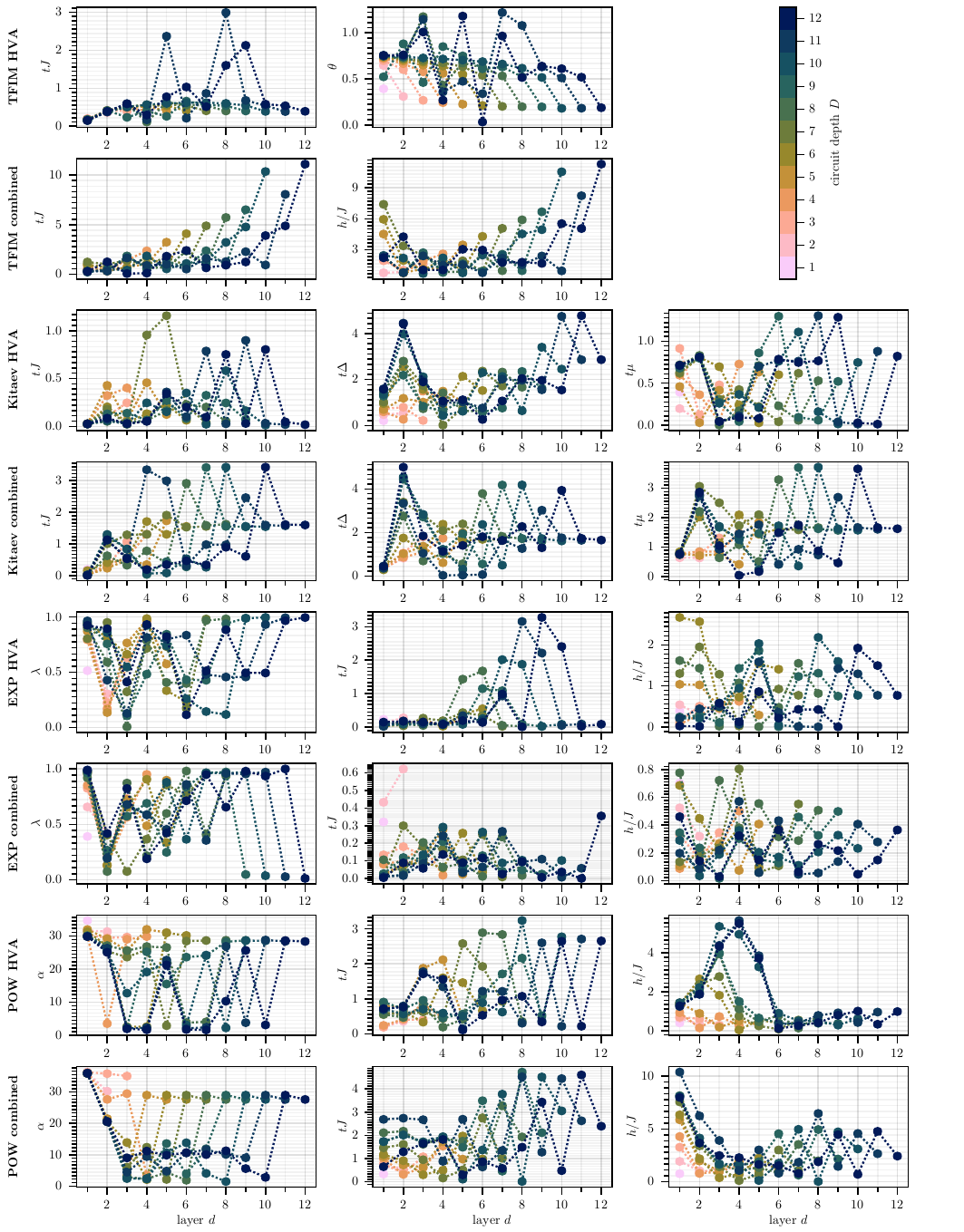}
 \caption{\label{fig:params}%
\textbf{Optimized variational parameters after natural-gradient optimization.} Optimization parameters for the representative system size $L=128a$, with colors indicating the total circuit depth $D$ and the horizontal axis showing the layer index $d\in[1,\ldots,D]$. The rows correspond to the TFIM HVA (\cref{eq:ansatz-NN-sep}), TFIM combined (\cref{eq:ansatz-NN-com}), exponential long-range (EXP, $\eta=\lambda$; \cref{eq:ansatz-LR-com}), power-law long-range (POW, $\eta=\alpha$; \cref{eq:ansatz-LR-com}), Kitaev HVA (\cref{eq:ansatz-Kitaev-HVA}), and Kitaev combined (\cref{eq:ansatz-Kitaev-com}) ansätze. Across all architectures, the optimized parameters remain bounded as the circuit depth increases, supporting the use of the circuit depth $D$ as the relevant resource measure in the scaling analysis.
}
\end{figure*}

As an illustration of the optimization procedure described above, Fig.~\ref{fig:params} shows the optimized variational parameters obtained after convergence for all ansätze at the representative system size $L=128a$. Beyond providing a concrete example of the parameter profiles produced by the natural-gradient optimization, these results also support an important methodological assumption of this work. Across all ansätze, the optimized parameters remain bounded as the circuit depth increases, rather than growing systematically with $D$.
This indicates that the increased expressiveness of the variational state is primarily achieved through the addition of layers, rather than through increasingly large parameter values, supporting the use of the circuit depth $D$ as the relevant measure of computational resources throughout the scaling analysis.

\section{Conformal perturbation theory modeling the VQE state}\label{app:scaling-CFT}

This appendix derives the effective scaling forms used to analyze the numerical results of Section~\ref{subsec:numerical}. Following the scaling hypothesis described in Section~\ref{sec:framework}, we model a finite-depth variational state as the ground state of an effective Hamiltonian
\begin{align}
    H = H_T + gV,
\end{align}
where $H_T$ is the critical target Hamiltonian and $V=\int_0^L\Phi(x)\,\dd x$ is the leading symmetry-allowed perturbation. This effective description predicts both the exponential cutoff of correlation functions used to extract the finite-depth correlation length $\xi_D$, and the finite-depth scaling of the variational energy. Throughout this appendix, $\ket{\psi}$ denotes the optimized variational state, $\ket{\Omega}$ the ground state of $H_T$, and $E_0(L)$ its finite-size ground-state energy.

\subsection{Correlation functions}\label{app:correlator}

We first derive the observable used to extract the finite-depth correlation length, namely the longitudinal spin--spin correlation function
\begin{align}
    C_{xx}(m,n)=\expval{\sigma^x_m\sigma^x_n}.
\end{align}
For the transverse-field Ising chain, the Jordan--Wigner transformation maps the spin operators to Majorana strings,
\begin{align}
   \sigma^x_m \sigma^x_n
   &=
   B_m A_{m+1} B_{m+1} A_{m+2}\cdots B_{n-1}A_n,
   \\
   A_l &= c_l+c_l^\dagger,\qquad
   B_l=c_l-c_l^\dagger,
\end{align}
for $m<n$~\cite{kitaev_unpaired_2001}. For a Gaussian (free-fermion) state, completely characterized by its correlation matrix $\Gamma$, Wick's theorem reduces the expectation value of this Majorana string to a Pfaffian which, for the present block structure, simplifies to the determinant of the $\abs{n-m}\times \abs{n-m}$ matrix of elementary contractions,
\begin{align}
   f(k,l)
   &\coloneqq
   \expval{B_kA_l}
   \nonumber\\
   &=
   \Gamma_{k+N,l}
   -\Gamma_{k,l}
   +\Gamma_{k+N,l+N}
   -\Gamma_{k,l+N},
\end{align}
where the four terms correspond to the correlators
$\expval*{c_k c_l}$,
$\expval*{c_k^\dagger c_l}$,
$\expval*{c_k c_l^\dagger}$,
and
$\expval*{c_k^\dagger c_l^\dagger}$.
The lattice correlator therefore becomes
\begin{align}
   C_{xx}(m,n)
   =
   \det[
   f(m+j-1,m+l)
   ]_{j,l=1}^{\abs{n-m}},
   \label{eq:cxx-det}
\end{align}
with the convention $C_{xx}(m,m)=1$.
Throughout our numerical calculations, one operator is fixed at the center of the chain while the second is displaced by a distance $r=\abs{m-n}$. Equation~\eqref{eq:cxx-det} therefore provides the exact lattice correlator against which the continuum scaling form is compared.

At criticality, the lattice two-point correlator approaches the conformal field theory prediction for the Ising order parameter, whose scaling dimension is $\Delta_\sigma=1/8$.
Given one field anchored on the center site and the other a distance $r$ away with $r < L$, the scaling form is given by
\begin{align}
   C_{xx}^{\infty}(r)
   \simeq
   \mathcal Z_\sigma
   \left[
      \frac{L}{\pi}
      \sin(\frac{\pi r}{L})
   \right]^{-2\Delta_\sigma},
   \label{eq:cxx-cft}
\end{align}
where the chord distance
\begin{align}
d_{\mathrm{chord}}(r,L)
=
\frac{L}{\pi}
\sin(\frac{\pi r}{L})
\end{align}
implements the conformal mapping from the finite interval to the bulk power law, reducing to $r$ when $r\ll L$, recovering $C_{xx} \propto r^{-1/4}$.

A finite variational depth introduces an infrared cutoff through the emergent correlation length $\xi_D$. Accordingly, we model the lattice correlator by multiplying the critical conformal prediction by an exponential envelope,
\begin{align}
   C_{xx}^{D}(r)
   \simeq
   \mathcal Z_\sigma
   \left[
      \frac{L}{\pi}
      \sin(\frac{\pi r}{L})
   \right]^{-2\Delta_\sigma}
   \exp(-r/\xi_D),
   \label{eq:cxx-finiteD}
\end{align}
valid within the scaling window
\begin{align}
a_{XX} \ll r \ll L \quad \text{and} \quad r/\xi_D \gtrsim \order{1},
\end{align}
where $a_{XX}$ denotes a non-universal ultraviolet cutoff below which lattice corrections become important. The exponential envelope is the observable manifestation of the effective relevant perturbation introduced in the main text: finite variational resources generate a finite infrared length scale $\xi_D$ exactly as a physical perturbation away from criticality would. Consequently, multiplying the measured correlator by the appropriate power of the chord distance removes the universal conformal power law and isolates the exponential decay used in Sec.~\ref{subsec:numerical} to extract $\xi_D$. The ultraviolet cutoff only affects the non-universal amplitude and short-distance corrections, and therefore does not modify the extracted correlation length.

\subsection{Energy shift}\label{app:energy-shift}

We now derive the finite-depth scaling of the variational energy error used in the main text. The observable of interest in this section is the absolute variational energy error,
\begin{align}
\delta E_{\rm abs}
=
\expval{H_T}{\psi}-E_0(L),
\end{align}
where $E_0(L)$ is the exact finite-size ground-state energy of the target Hamiltonian.

Because $\ket{\psi}$ is a variational optimum, the energy is stationary with respect to infinitesimal changes of the perturbation strength. Consequently, the linear correction vanishes,
\begin{align}
\eval{\pdv{\delta E_{\rm abs}}{g}}_{g=0}
=
0,
\end{align}
and the leading contribution is quadratic,
\begin{align}
\delta E_{\rm abs}
=
g^2\chi_m+\order{g^3},
\end{align}
where
\begin{align}
\chi_m
=
\sum_{n>0}
\frac{\abs{\mel{n}{V}{\Omega}}^2}
{E_n-E_0}
\end{align}
is the generalized susceptibility associated with the perturbation $V$.

Using the spectral representation,
\begin{align}
\frac1{E_n-E_0}
=
\int_0^\infty
e^{-(E_n-E_0)\tau}
\,\dd\tau,
\end{align}
the susceptibility becomes an integrated Euclidean correlation function,
\begin{align}
\chi_m
=
\int_0^\infty
\expval{W(\tau)W(0)}_C
\,\dd\tau,
\end{align}
with
\begin{align}
W(\tau)
=
e^{(H_T-E_0)\tau}
V
e^{-(H_T-E_0)\tau}.
\end{align}
The perturbation $V$ contains a spatial integral of the local scaling operator $\Phi$. In order to progress, we compute the leading bulk contribution by performing the spatial integration explicitly, which would have been possible with true translational invariance,
\begin{align}
\chi_m
\simeq
L
\int
G(u,\tau)
\,\dd u\,\dd\tau,
\end{align}
where
\begin{align}
G(u,\tau)
=
\expval{\Phi(u,\tau)\Phi(0,0)}_C.
\end{align}

Within conformal perturbation theory, the connected correlator is approximated by its continuum critical form,
\begin{align}
G(r)
\propto
r^{-2\Delta_\Phi},
\end{align}
between a non-universal ultraviolet cutoff $a_E$ and an infrared cutoff
\begin{align}
R\simeq\min(L,\xi).
\end{align}
The susceptibility therefore scales as
\begin{align}
\chi_m
&\propto
L
\int_{a_E}^{R}
r^{1-2\Delta_\Phi}
\,\dd r.
\end{align}
For a relevant perturbation with $\Delta_\Phi \leq 1$,
\begin{align}
\chi_m
\propto
L
\frac{
R^{2-2\Delta_\Phi}
-
a_E^{2-2\Delta_\Phi}}
{2-2\Delta_\Phi},
\end{align}
while the borderline case $\Delta_\Phi=1$ gives the logarithmic behavior
\begin{align}
\chi_m
\propto
L
\ln(R/a_E).
\end{align}

In the finite-depth regime, $\xi\ll L$, the infrared cutoff is set by the correlation length, $R=\xi$. Since the running coupling is of order unity at this scale, the leading contribution becomes
\begin{align}
L\,
\delta E_{\rm abs}
\propto
\left(\frac{L}{\xi}\right)^2,
\end{align}
up to the logarithmic factor
\begin{align}
\ln(\xi/a_E)
\end{align}
for the marginal Ising energy operator with $\Delta_\Phi=1$. This is the scaling form used throughout the main text.

The previous result describes the scaling of the total energy correction. For the numerical analysis it is more convenient to isolate the bulk contribution by subtracting the known finite-size CFT correction of the critical Hamiltonian. We therefore define
\begin{align}
\delta e_{\mathrm{bulk}}
=
\frac{E(L,D)}{L}
-
e_0
-
\frac{e_{\mathrm{boundary}}}{L},
\end{align}
where $e_0$ and $e_{\mathrm{boundary}}$ are the bulk and boundary energy densities of the target model. Using the finite-size expansion of the exact critical ground-state energy,
\begin{align}
E_0(L)
=
Le_0
+
e_{\mathrm{boundary}}
-
\frac{\pi cv}{24L}
+
\order{L^{-2}},
\end{align}
one obtains
\begin{align}
L^2\delta e_{\mathrm{bulk}}
=
L\,\delta E_{\mathrm{abs}}
-
\frac{\pi cv}{24}
+
\order{L^{-1}}.
\end{align}
In the finite-depth regime, inserting the conformal perturbation result above gives
\begin{align}
L^2\delta e_{\mathrm{bulk}}
\simeq
B
\left(\frac{L}{\xi}\right)^2
\ln(\frac{\xi}{a_E})
-
\frac{\pi cv}{24},
\label{eq:bulk-energy-shift-FDS}
\end{align}
where $B$ is a non-universal amplitude. Away from the marginal case, the logarithmic factor is absent.

\subsubsection{Jackknife uncertainty estimate}\label{par:jackknife}

To quantify the stability of the extracted exponent with respect to the limited set of available circuit depths, we use a delete-one-depth jackknife resampling~\cite{wu_jackknife_1986}:
For $n_D$ fitted depths, we repeat the complete fit $n_D$ times, each time omitting all data associated with one depth. If $\kappa_{(-i)}$ denotes the estimate obtained after removing depth $D_i$ and $\overline{\kappa}_{(-)}=n_D^{-1}\sum_i\kappa_{(-i)}$, the reported jackknife standard error is
\begin{align}
 \sigma_{\mathrm{JK}}
 =
 \sqrt{\frac{n_D-1}{n_D}
 \sum_{i=1}^{n_D}
 \left(\kappa_{(-i)}-\overline{\kappa}_{(-)}\right)^2}.
\end{align}
We jackknife resample over depths $D$, rather than individual $(L,D)$ points, because the scaling exponent $\kappa$ is determined by the variation with $D$ and all system sizes at a given depth share the same leverage on that scaling.
In contrast, the standard error obtained from the local curvature of a single least-squares fit is conditional on the selected data and scaling model and can therefore be misleadingly small when only a few depths constrain the exponent.
In the present data, the fitted exponent can change appreciably when one depth is removed.
Consequently, the jackknife error is the more relevant measure because it exposes this sensitivity directly.
It should, however, be interpreted as a diagnostic on fit stability rather than a complete statistical or systematic uncertainty. In particular, when only a few depths are available, neither a small formal fit error nor the jackknife standard error establishes that the asymptotic scaling regime has been reached.
Instead, strong variation under leave-one-depth-out resampling signals that additional depths or larger system sizes are required for a well-grounded estimate.

\subsubsection{Variable projection fit with the logarithmic correction}\label{par:varpro}

Using the finite-depth scaling hypothesis,
\begin{align}
    \xi = A_\xi D^\kappa,
\end{align}
the energy-shift scaling derived above becomes
\begin{align}
   y(L,D)
   &\coloneqq
   L\,\delta E_\mathrm{abs}(L,D)
   \nonumber\\
   &=
   P\,L^2D^{-2\kappa}
   \left(\kappa\ln D + Q\right),
   \label{eq:varpro-model}
\end{align}
where $P\propto B/A_\xi^2$ is a non-universal amplitude and
\begin{align}
Q=\ln(A_\xi/a_E)
\end{align}
contains the ultraviolet cutoff introduced by the continuum approximation.

For fixed values of $\kappa$ and $Q$, Eq.~\eqref{eq:varpro-model} is linear in the amplitude $P$. We therefore use a variable-projection procedure~\cite{golub_differentiation_1973}: at each trial value of $\kappa$, we scan $Q$ over its physically allowed interval and determine the optimal $P$ analytically by weighted linear least squares. The residual minimized by the outer one-dimensional optimization over $\kappa$ is thus already profiled over both $P$ and $Q$. To avoid the largest system sizes dominating the fit through the overall $L^2$ prefactor, we use relative least-squares weights $w\propto y^{-2}$.

A direct nonlinear fit of all parameters is numerically unstable because the logarithmic term $\kappa\ln D$ and the constant $Q$ become nearly collinear over the accessible depth window. Consequently, changes in $Q$ can partially compensate changes in $\kappa$, leading to poorly conditioned fits. We regularize this degeneracy by constraining $Q$ through its physical interpretation as the ultraviolet cutoff. Since the cutoff satisfies
\begin{align}
1\le a_E\le \xi(D_{\min}),
\end{align}
the allowed range becomes
\begin{align}
-\kappa\ln D_{\min}
\le
Q
\le
\ln A_\xi,
\end{align}
where both $A_\xi$ and $D_{\min}$ are obtained independently from the correlation-length analysis of Appendix~\ref{app:correlator}. For every admissible $Q$ at fixed $\kappa$, the amplitude $P$ is projected out analytically; the value of $Q$ yielding the smallest weighted residual defines the profiled residual used in the outer optimization over $\kappa$. This procedure stabilizes the extraction of the scaling exponent, while the fitted cutoff itself remains comparatively uncertain because of the residual correlation between $Q$ and $\kappa$. Once the fit is completed, the corresponding ultraviolet cutoff is recovered from
\begin{align}
a_E=A_\xi e^{-Q}.
\end{align}

\section{Complete scaling analysis for all variational ansätze}\label{app:detailed-plots}

For the TFIM HVA, the complete scaling analysis is presented in the main text. Here we collect the corresponding analyses for all remaining variational ansätze considered in this work. Each figure follows the same workflow: extraction of the finite-depth correlation length from the spin correlator, determination of the scaling exponent from the energy-collapse analysis, and quasiparticle occupations in the eigenbasis of the target Hamiltonian. These data underlie the scaling exponents summarized in Table~\ref{tab:kappas} and enable a direct comparison of the different variational architectures at the level of the raw $(L,D)$ data.

For the TFIM combined and Kitaev combined ansätze, the shallowest circuit depths do not yet lie in the asymptotic finite-depth scaling regime. This is apparent from the energy-collapse analysis, where these data do not approach the same asymptotic plateau as the deeper circuits. Accordingly, the $D=1$ data for the TFIM combined ansatz and the $D=1,2$ data for the Kitaev combined ansatz are excluded from the extraction of the scaling exponent reported in Table~\ref{tab:kappas}.

% \clearpage

\begin{figure*}
  \centering
  \includegraphics[width=\textwidth]{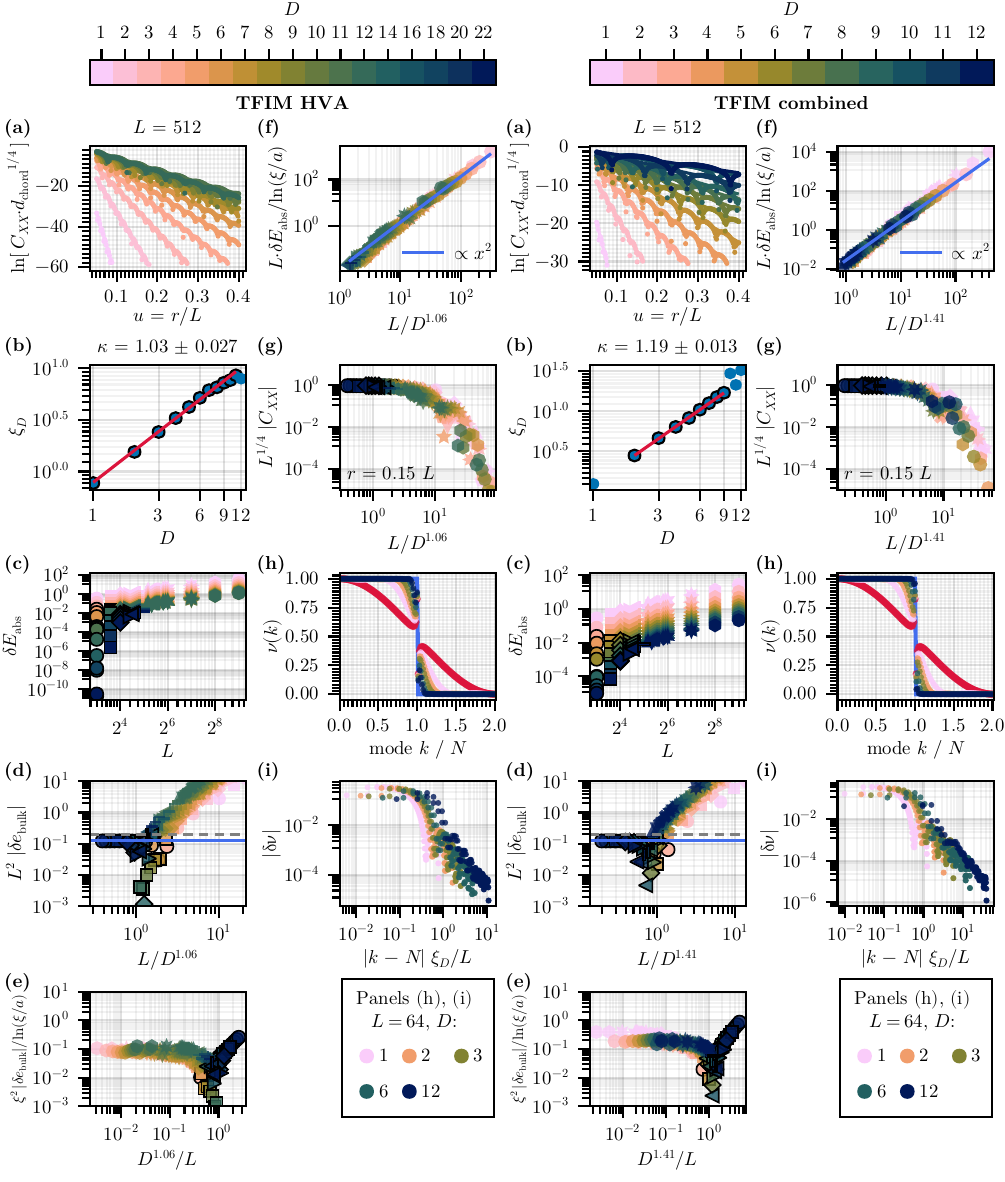}
  \caption{\label{fig:mega_tfim}
\textbf{Complete scaling analysis for the nearest-neighbor TFIM variational ansätze.}
Columns correspond to the TFIM HVA~\eqref{eq:ansatz-NN-sep}, TFIM combined~\eqref{eq:ansatz-NN-com} from left to right. Throughout, the circuit depth $D$ is color-coded according to the corresponding color bar, while different marker shapes denote different system sizes $L$.
(a) Chord-corrected spin--spin correlation function; markers show the numerical data and solid lines the exponential fits to the extracted envelope.
(b) Correlation length $\xi_D$ as a function of circuit depth together with the algebraic fit $\xi_D=A_\xi D^\kappa$; the shaded region indicates the fitted standard error.
(c) Absolute variational energy error $\delta E_\mathrm{abs}(L,D)$ versus system size.
(d,e) Complementary views of the bulk energy density shift: (d) $L^2\abs{\delta e_\mathrm{bulk}}$ versus $L/\xi_D$, highlighting the finite-size regime; (e) $\xi_D^2\abs{\delta e_\mathrm{bulk}}/\ln(\xi_D/a)$ versus $\xi_D/L$, highlighting the finite-depth regime, where the data approach a predicted plateau.
(f) Energy-collapse analysis obtained from the VARPRO fit as a function of $L/\xi_D$; the inset shows the corresponding collapse of the chord-corrected correlation function at fixed $u=r/L$.
(g) Collapse of the correlation function $C_{XX}(u,L/\xi)$ vs $ x\propto L/ \xi_D$ for fixed $u=0.15$.
(h,i) Quasiparticle occupations in the eigenbasis of the target Hamiltonian as a function of the relative mode index $k/N$. The solid blue line denotes the exact ground-state occupation, while (i) shows the deviation $\abs{\delta\nu_k}$ as a function of the rescaled distance from the Fermi surface, $\abs{k-N}\xi_D/L$.
}
\end{figure*}

\begin{figure*}
  \centering
  \includegraphics[width=\textwidth]{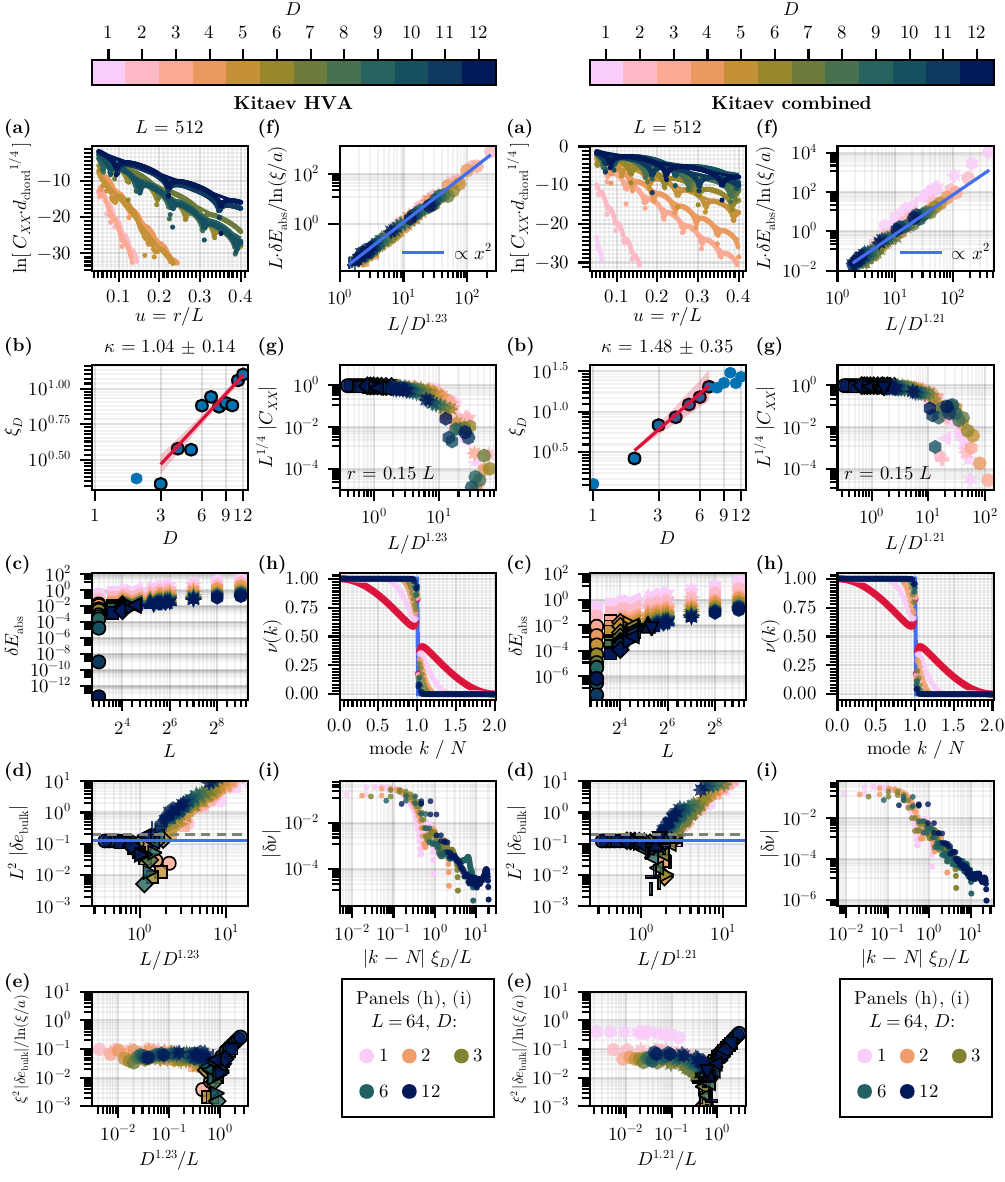}
  \caption{\label{fig:mega_kitaev}
\textbf{Complete scaling analysis for the nearest-neighbor Kitaev variational ansätze.}
Same as Fig.~\ref{fig:mega_tfim} but for the Kitaev HVA~\eqref{eq:ansatz-Kitaev-HVA}, and Kitaev combined~\eqref{eq:ansatz-Kitaev-com} shown from left to right.
}
\end{figure*}

\begin{figure*}
  \centering
  \includegraphics[width=\textwidth]{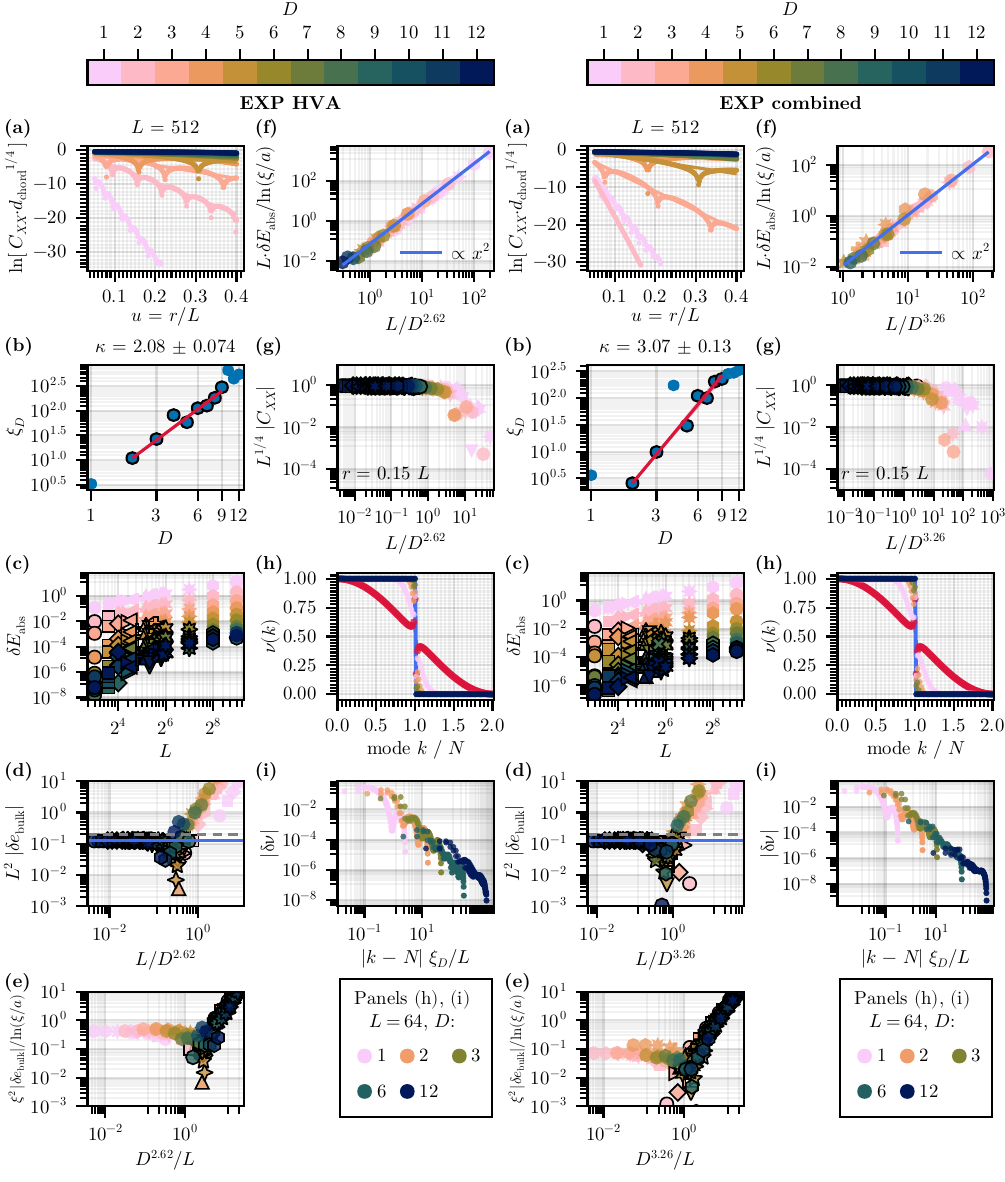}
  \caption{\label{fig:mega_exp}
\textbf{Complete scaling analysis for the exponentially long-range variational ansätze.}
Same as Fig.~\ref{fig:mega_tfim}, but for the EXP HVA~\eqref{eq:ansatz-LR-sep} with $\eta=\lambda$, EXP combined~\eqref{eq:ansatz-LR-com} with $\eta=\lambda$, shown from left to right.
}
\end{figure*}

\begin{figure*}
  \centering
  \includegraphics[width=\textwidth]{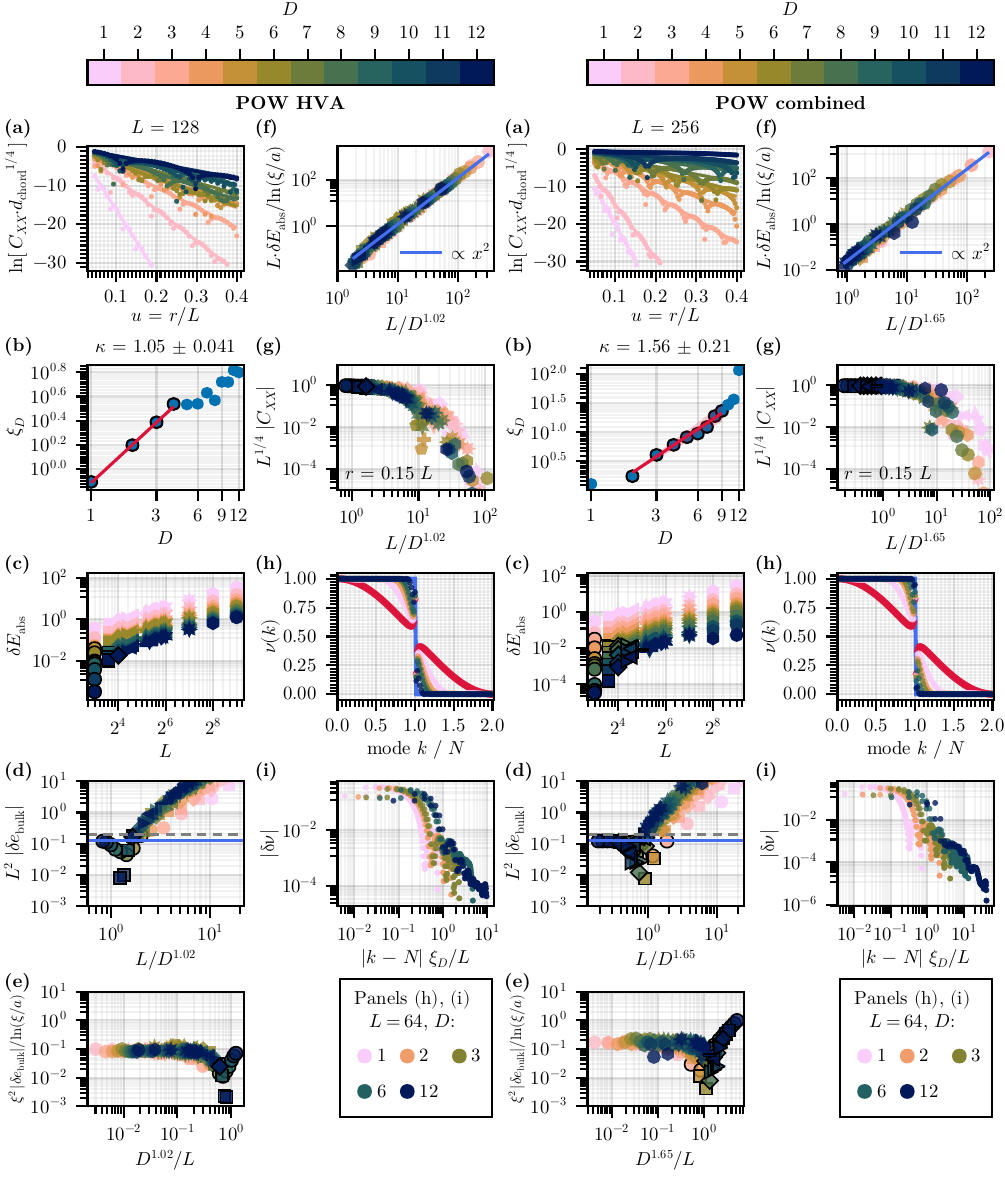}
\caption{\label{fig:mega_pow}%
\textbf{Complete scaling analysis for the power-law long-range variational ansätze.}
Same as Fig.~\ref{fig:mega_tfim}, but for the POW HVA~\eqref{eq:ansatz-LR-sep} with $\eta=\alpha$, and POW combined~\eqref{eq:ansatz-LR-com} with $\eta=\alpha$, shown from left to right.
}
\end{figure*}

\end{document}